  \documentclass[multphys,vecphys,numreferences]{svmult}

\usepackage{makeidx}
\usepackage{multicol}
\usepackage{footmisc}
\usepackage{url}
\usepackage{cite}

 \newcommand{\etal}{ {\sl et~al}.}           %%%  from macr.tex
 \newcommand{\abinitio}{{\sl ab~initio} }

\usepackage{dcolumn} %%% Align table columns on decimal point: not work always
\usepackage{amsmath}
\usepackage[dvips]{graphicx}
\usepackage{bm}
\begin{document}

 \newcommand{\ande}{ {\&} }
 \newcommand{\etale}{ {\sl et~al.}}
 \newcommand{\eref}[1]{(\ref{#1})}
 \newcommand{\eeref}[1]{eq.~(\ref{#1})}
 \newcommand{\erefs}[1]{eqs.~(\ref{#1})}
 \newcommand{\eerefs}[1]{eqs.~(\ref{#1})}
 \newcommand{\frefs}[1]{figures~\ref{#1}}
 \newcommand{\tref}[1]{table~\ref{#1}}
 \newcommand{\trefs}[1]{tables~\ref{#1}}
 \newcommand{\Eref}[1]{Eq.~(\ref{#1})}
 \newcommand{\Erefs}[1]{Eqs.~(\ref{#1})}
 \newcommand{\Frefs}[1]{Figures~\ref{#1}}
 \newcommand{\Tref}[1]{Table~\ref{#1}}
 \newcommand{\Trefs}[1]{Tables~\ref{#1}}
 \newcommand{\Vres}{\ensuremath{V'_{e,e}}}
 \newcommand{\veps}{\ensuremath{\varepsilon}}
 \newcommand{\Aeff}{\ensuremath{A_{\rm eff}}}
 \newcommand{\cm}{\mbox{cm}\ensuremath{^{-1}}} %%% English
 \newcommand{\rtw}{\rightarrow}
 \newcommand{\half}{\ensuremath{\frac{1}{2}}}
 \newcommand{\un}[1]{\underline{\vphantom{pP}#1}}
 \newcommand{\unem}[1]{\underline{\em\vphantom{pP}#1}}
 \newcommand{\Hzecm}{\ensuremath{\frac{\mbox{\normalsize Hz}}
                                      {\mbox{\normalsize e~cm}}}}
 \newcommand{\eHzecm}{\ensuremath{\frac{\mbox{\normalsize Hz}}
                                      {\mbox{\normalsize e~cm}}}}
 \newcommand{\opH}{{\bf H}}
 \newcommand{\opHEf}{\opH^{\rm Ef}}
 \newcommand{\opHeff}{\opH_{\rm eff}}
 \newcommand{\opHso}{\opH^{\rm SO}}
 \newcommand{\opHSO}{\opH^{\rm SO}}
 \newcommand{\opHo}{\opH^{[0]}} 
 \newcommand{\opHFr}{\opH^{\rm Fr}}
 \newcommand{\opHfr}{\opH^{\rm Fr}}
 \newcommand{\oph}{{\bf h}}
 \newcommand{\opU}{{\bf U}}
 \newcommand{\opUEf}{\opU^{\rm Ef}}
 \newcommand{\opUef}{\opU^{\rm Ef}}
 \newcommand{\opUefHuz}{\opU^{\rm Ef}_{\rm Huz}}
 \newcommand{\opUSfC}{\ensuremath{\ensuremath{\opU}^{\rm SfC}}}
 \newcommand{\opW}{{\bf W}}
 \newcommand{\opF}{{\bf F}}
 \newcommand{\opS}{{\bf S}}
 \newcommand{\oprn}{\vec{r}_{n}}
 \newcommand{\sigp}{(\vec{\sigma}{\cdot}\vec{\bf p})}
 \newcommand{\alphp}{(\vec{\alpha}{\cdot}\vec{\bf p})}
 \newcommand{\opV}{{\bf V}}
 \newcommand{\opVr}{{\it V(r)}}     % for s2comp
 \newcommand{\opVnuc}{{\opV}}
 \newcommand{\opVext}{\opV^{\rm ext}}
 \newcommand{\opVcor}{\opV^{\rm corr}}
 \newcommand{\opVcorr}{\opV^{\rm corr}}
 \newcommand{\opX}{{\bf X}}
 \newcommand{\opJ}{{\bf J}}
 \newcommand{\opK}{{\bf K}}
 \newcommand{\oppsJ}{\widetilde{\bf J}}
 \newcommand{\oppsK}{\widetilde{\bf K}}
 \newcommand{\opP}{{\bf P}}
 \newcommand{\opPl}{\opP_l}
 \newcommand{\opPlj}{\opP_{l\pm}}
 \newcommand{\vecr}{\vec{\it r}}
 \newcommand{\vecp}{\vec{\bf p}}
 \newcommand{\vecalp}{\vec{\alpha}}
 \newcommand{\vecsig}{\vec{\sigma}}
 \newcommand{\OCin}{\mbox{֕}\ensuremath{^<}}
 \newcommand{\OMC}{\ensuremath{\mbox{֕}^>}}
 \newcommand{\opNomc}{\ensuremath{{\bf N}^>}}
 \newcommand{\Nomc}{\ensuremath{N^>}}
 \newcommand{\DNomc}{\ensuremath{\Delta N}}
 \newcommand{\Rc}{\ensuremath{R_{\rm c}}}
 \newcommand{\Romc}{\ensuremath{r_{\rm omc}}}
 \newcommand{\Rv}{\ensuremath{r_{\rm v}}}
 \newcommand{\Rrest}{\ensuremath{R_C}}
 \newcommand{\Eco}{\ensuremath{E_{\rm core}}}
 \newcommand{\EcoNi}{\ensuremath{\Eco^{N_i}}}
 \newcommand{\dEcoNiDN}{\ensuremath{\delta\EcoNi(\DNomc_i)}}
 \newcommand{\Efr}{\ensuremath{E_{\rm fr}}}
 \newcommand{\dEfr}{\ensuremath{\delta\Efr}}
 \newcommand{\dPsi}{\ensuremath{\delta\Psi}}
 \newcommand{\Psifr}{\ensuremath{\Psi_{\rm fr}}}
 \newcommand{\dPsifr}{\ensuremath{\delta\Psifr}}
 \newcommand{\Esm}{\ensuremath{E_{\rm sm}}}
 \newcommand{\dEsm}{\ensuremath{\delta\Esm}}
 \newcommand{\Eef}{\E_{\rm eff}}

 \newcommand{\DCBrev}{Schwerdtfeger:02bb,Schwerdtfeger:04aa,Hirao:04}
 \newcommand{\Kutzeln}{Kutzelnigg:84,Kutzelnigg:89,Kutzelnigg:90,Kutzelnigg:02}
 \newcommand{\KinBal}{Lee:82,Schwarz:82,Schwarz:82b,Dyall:84c}
 \newcommand{\TupTwoAt}{Tupitsyn:00}
 \newcommand{\TwoCompAppr}{Wood:78,Barthelat:80,Lenthe:93,Wolf:02}
 \newcommand{\RCCrev}{Kaldor:99,Landau:01c,Visscher:01}
 \newcommand{\Nether}{Visscher:96bb,DeJong:96,DeJong:99}
 \newcommand{\TAU}{Eliav:96,Eliav:98,Kaldor:99,Landau:01c}
 \newcommand{\SODCIp}{Buenker:99,Alekseyev:04,Titov:01}
 \newcommand{\Christ}{Christiansen:79}
 \newcommand{\cErmler}{Ermler:78}
 \newcommand{\cKahn}{Kahn:76}
 \newcommand{\CommentA}{Mosyagin:98cd}
 \newcommand{\MosyaginA}{Mosyagin:94}
 \newcommand{\MosyaginB}{Mosyagin:97}
 \newcommand{\MosyaginHg}{Mosyagin:98bd,Mosyagin:00,Mosyagin:01b,Mosyagin:02}
 \newcommand{\TitovA}{Titov:91}
 \newcommand{\GRECP}{Titov:99,Titov:00a,Titov:02Dism,Petrov:04b,Mosyagin:04a}
 \newcommand{\GRThGr}{Titov:99}
 \newcommand{\RECPcomp}{Tupitsyn:95,Mosyagin:97,Titov:99,Titov:00rd,%
Mosyagin:00,Isaev:00,Titov:01,Mosyagin:01b}
 \newcommand{\TupitsynA}{Tupitsyn:95}
 \newcommand{\HFD}{Bratzev:77rd}
 \newcommand{\TupitsynB}{Bratzev:77rd}
 \newcommand{\cBuenker}{Buenker:74,Buenker:75,Buenker:78,Alekseyev:94}
 \newcommand{\PPsolids}{Hamann:79,Bachelet:82,Bachelet:82b,Cohen:83}
 \newcommand{\EAPP}{Kuchle:91,Dolg:92,Haussermann:93,Dolg:93}
 \newcommand{\SeparablePP}{Blochl:90,Vanderbilt:90,Theurich:01}
 \newcommand{\PPsolidsRev}{Heine:73rd,Bachelet:82,Cohen:83,Abarenkov:96rd,%
Hartwigsen:98}
 \newcommand{\CPP}{Fuentealba:83b,Mueller:84a,Mueller:84b,Stoll:84,%
Kuchle:91,Foucrault:92,Magnier:93,Leininger:97}
 \newcommand{\HuzinagaPP}{Bonifacic:74,Katsuki:88b,Seijo:04}
 \newcommand{\RECPrev}{Ermler:88,Schwerdtfeger:03,YSLee:04}
 \newcommand{\SemiLocRECP}{Ermler:88,Schwerdtfeger:03,YSLee:04,Teichteil:04}
 \newcommand{\ECPhistB}{Phillips:59,Abarenkov:65,Weeks:68,Goddard:68,%
Durand:75,Kahn:76,Lee:77,Christiansen:79,Hamann:79,Hafner:79}
 \newcommand{\OCR}{Phillips:59,Pacios:85,Titov:85Dism,Titov:92A,Titov:96,%
Blochl:94}
 \newcommand{\PToddMolCalc}{Kozlov:87,Dmitriev:92,Titov:96b,Kozlov:97,%
Mosyagin:98,Petrov:02,Isaev:04,Petrov:05a}
 \newcommand{\PNCrev}{Commins:99,Sapirstein:02aa,Berger:04,Ginges:04,Erler:04}
 \newcommand{\BSSE}{Gutowski:86,Liu:89}
 \newcommand{\GCbasis}{Mosyagin:98bd,Mosyagin:00,Isaev:00,Mosyagin:01b}
 \newcommand{\Tlcalc}{Rakowitz:96,Wahlgren:97,Leininger:97}
 \newcommand{\cHFJ}{HFJ,Tupitsyn:95}
 \newcommand{\cHFDB}{HFDB,Bratzev:77,Tupitsyn:02A}
 \newcommand{\pMOLGEP}{{\sc molgep}}
 \newcommand{\pHFD}{{\sc hfd}}
 \newcommand{\pHFDB}{{\sc hfdb}}
 \newcommand{\pHFDCI}{{\sc hfd/ci}}
 \newcommand{\pHFJ}{{\sc grecp/hfj}}
 \newcommand{\pMOLCAS}{{\sc molcas}}
 \newcommand{\pCCSDT}{{\sc cc-sdt}}
 \newcommand{\pRCC}{{\sc rcc}}
 \newcommand{\pRCCSD}{{\sc rcc-sd}}
 \newcommand{\pRelCCSDT}{{\sc relccsd(t)}}
 \newcommand{\pMRDCI}{{\sc mrd-ci}}
 \newcommand{\pANOL}{{\sc ano-l}}
 \newcommand{\pPTdCI}{{\sc pt2/ci}}

%%\include{macr-nam}        %%%     Names of researches translated to Russian

%tav
% \title*{Study of Parity Violation Effects in Polar Heavy-Atom Molecules}
  \title*{Study of P,T-Parity Violation Effects in Polar Heavy-Atom Molecules}
%end

 \author{A.V.\ Titov\inst{1}\and N.S.\ Mosyagin\inst{1}\and
 A.N.\ Petrov\inst{1}\and T.A.\ Isaev\inst{1}\and D.\ DeMille\inst{2}}

%\affiliation{}
\institute{Petersburg Nuclear Physics Institute, 
           Gatchina, St.-Petersburg district 188300, Russia;
           \texttt{titov@pnpi.spb.ru; http://www.qchem.pnpi.spb.ru} \and
           Physics Department, Yale University, New Haven,
            Connecticut 06520, USA}

%\titlerunning{PNC effects in molecules}
%\authorrunning{A.V.\ Titov \etal}

\maketitle
%tav_ ??? temporarily added:
% \marginpar{\vspace{-55mm}Prog.Theor.Chem.Phys., 2005, in press.}
%end

\date{\today}

\begin{abstract}

 Investigation of P,T-parity nonconservation (PNC) phenomena is of 
 fundamental importance for physics.  Experiments
%tav2 on
      to
 search for PNC
 effects have been performed on TlF and YbF molecules and are in 
 progress for PbO and PbF molecules.  For interpretation of molecular
 PNC experiments it is necessary to calculate those needed molecular
 properties which cannot be measured.  In particular, electronic densities in
 heavy-atom cores are required for interpretation of the measured data in terms of the
 P,T-odd properties of elementary particles or P,T-odd interactions between
 them.  Reliable calculations of the core properties (PNC effect, hyperfine
 structure etc., which are described by the operators heavily concentrated in
 atomic cores or on nuclei) usually require accurate accounting for both
 relativistic and correlation effects in heavy-atom systems.  In this paper,
 some basic aspects of the experimental search for PNC effects in
 heavy-atom molecules and the computational methods used in their electronic 
 structure calculations are discussed.  The latter include the generalized 
%tav
% RECP
  relativistic effective core potential (GRECP)
%end 
 approach and the methods of nonvariational and variational
 one-center restoration of correct shapes of four-component spinors in atomic
 cores after a two-component 
%tav 
% RECP
  GRECP
%end
 calculation of a molecule.  Their efficiency
 is illustrated with calculations of parameters of the effective P,T-odd
 spin-rotational Hamiltonians in the molecules PbF, HgF, YbF, BaF, TlF, and PbO.
\end{abstract}

%         nonconsevation.}
%         atom in a molecule, one-center restoration.}

%============================================================================
\section*{Introduction}
 \label{sIntro}
%============================================================================
%\vspace{20mm}

 It is well recognized that polar diatomics containing heavy elements are very
 promising objects for the experimental search for the break of inversion
 symmetry~(P) and time-reversal invariance~(T).  Though the search for the
 P,T-parity nonconservation (PNC) effects in heavy atoms and heavy-atom
 molecules has produced null results up to now, there are serious reasons to
 search for them with the presently accessible (expected) level of 
 experimental sensitivity.  The observation of non-zero P,T-odd effects at this
 level would indicate the presence of so-called ``new physics''\cite{Erler:04}
 beyond the Standard Model (SM) of electroweak and strong interactions
 \cite{Glashow:61, Weinberg:67, Salam:68, Weinberg:72} that is certainly of
 fundamental importance. Despite well known drawbacks and unresolved problems
 of the Standard Model (radiative corrections to the Higgs mass are
 quadratically divergent; rather artifiÓial Higgs mechanism of symmetry
 breaking is not yet verified in experiment; the problem of CP-violation is not
 well understood, where ``C'' is charge conjugation symmetry etc.) there are no
 experimental data available which would be in direct contradiction with this
 theory (see section \ref{sStatus} and papers \cite{Commins:99, Erler:04} for
 more details and references).  In turn, some popular extensions of the
 Standard Model, which allow one to overcome its disadvantages, are not
 confirmed experimentally.

 A crucial feature of PNC experiments in atoms, molecules, liquids or solids
 is that for interpretation of measured data in terms of fundamental constants
 of the P,T-odd interactions, one must calculate those properties of the
% many-electron
 systems, which establish a connection between the measured data and
 studied fundamental constants (see section \ref{sPTexp}).  These properties
 are described by operators heavily concentrated near or on heavy nuclei;
 they cannot be measured and their theoretical study is not a trivial task.
 During the last several years the significance of (and requirement for) \abinitio
 calculation of electronic structure providing a high level of reliability and
 accuracy in accounting for both relativistic and correlation effects has only
 increased (see sections \ref{sStatus} and \ref{sPbO}).

 The main goal of the paper is to discuss the present status of relativistic
 calculations of P,T-odd properties in heavy-atom molecules, the two-step
 methodology used in these calculations, and the accuracy of this method.
 The historical background of the PNC study in
 atoms and molecules, its current status and some general remarks on the PNC
 experiments are presented in sections \ref{sPThist}, \ref{sStatus} and
 \ref{sPTexp}, correspondingly.  The \abinitio relativistic methods and the
 two-step techniques of calculation designed for studying PNC properties in
 heavy-atom molecules are discussed in sections \ref{sHAMcalc} and
 \ref{s2step}.  The calculations of PNC properties and hyperfine structure in
 molecules PbF, HgF, YbF, BaF, TlF and PbO are presented in sections
 \ref{sPbFHgF}--\ref{sPbO}.  Concluding remarks are outlined in section
 \ref{sConcl}.

%============================================================================
\section{Study of P- and T-parity nonconservation effects in heavy-atom
         molecules:  Historical background}
 \label{sPThist}
%============================================================================

 After discovery of the combined charge and space parity violation, or
 CP-violation, in $K^0_L$-meson decay \cite{Christenson:64}, the search for the
 electric dipole moments (EDMs) of elementary particles has become one of the
 most fundamental problems in physics \cite{Commins:99, Sapirstein:02aa,
 Berger:04, Ginges:04, Erler:04}.  A permanent EDM is induced by the weak
 interaction that breaks both the space symmetry inversion and time-reversal
 invariance \cite{Landau:57}.  Considerable experimental effort has been
 invested in probing for atomic EDMs induced by EDMs of the proton, neutron and
 electron, and by P,T-odd interactions between them.  The best available
 restriction for the electron EDM, $d_e$, was obtained in the atomic Tl
 experiment \cite{Regan:02}, which established an upper limit of
 $|d_e|<1.6\times10^{-27}\ e\cdot$cm, where $e$ is the charge of the electron.
 The benchmark upper limit on a nuclear EDM is obtained in atomic experiment on
 $^{199}$Hg \cite{Romalis:01}, $|d_{\rm Hg}|<2.1\times 10^{-28}\ e\cdot$cm,
 from which the best restriction on the proton EDM, $|d_p|<5.4\times 10^{-24}\
 e\cdot$cm, was also recently obtained by Dmitriev\ande Sen'kov
 \cite{Dmitriev:03} (the previous upper limit on the proton EDM was obtained in
 the TlF experiment, see below).

 Since 1967, when Sandars suggested
%tav2 to use
     the use of
%end
 polar heavy-atom molecules in the
 experimental search for the proton EDM \cite{Sandars:67}, molecules have been
 considered the most promising objects for such experiments.  Sandars also
 noticed earlier \cite{Sandars:65} that the P- and P,T-parity nonconservation
 effects are strongly enhanced in heavy atoms due to relativistic and other
 effects.  For example, in paramagnetic atoms the enhancement factor for an
 electron EDM, $d_{\rm atom}/d_e$, is roughly proportional to $\alpha^2 Z^3
 \alpha_D$, where $\alpha \approx 1/137$ is the fine structure constant, $Z$ is
 the nuclear charge and $\alpha_D$ is the atomic polarisability.  It can be of
 order 100 or greater for highly polarizable heavy atoms ($Z\ge50$).
 Furthermore, the effective intramolecular electric field
%tav
% $E_{\rm mol}$
 acting on electrons in polar molecules can be
% a few
  five or more
 orders of magnitude higher than the maximal field
% $E_{\rm ext}$ accessible in a laboratory, $E_{\rm mol}/E_{\rm ext}\sim10^{5}$.
  accessible in a laboratory.
%end 
 The first molecular EDM experiment was performed
 on TlF by Sandars\etal\ \cite{Hinds:76} (Oxford, UK); it was interpreted as a
 search for the proton EDM and other nuclear P,T-odd effects.
 In 1991, in the last series of the $^{205}$TlF experiments by Hinds\etal\
 \cite{Cho:91} (Yale, USA), the restriction $d_p = (-4\pm6){\times}10^{-23}\
 e{\cdot}{\rm cm}$ was obtained (this was recalculated in 2002 by Petrov\etal\
 \cite{Petrov:02} as $d_p = (-1.7 \pm 2.8){\times}10^{-23}\ e{\cdot}{\rm cm}$).

 In 1978 the experimental investigation of the electron EDM and other PNC
 effects was further stimulated by Labzowsky\etal\
 \cite{Labzowsky:78,Gorshkov:79} and Sushkov\ande Flambaum \cite{Sushkov:78}
 who clarified the possibilities of additional enhancement of these effects in
 diatomic radicals like BiS and PbF due to the closeness of levels of opposite
 parity in $\Omega$-doublets having a $^2\Pi_{1/2}$ ground state.  Then
 Sushkov\etal\ \cite{Sushkov:84} and Flambaum\ande Khriplovich
 \cite{Flambaum:85b} suggested the use of $\Omega$-doubling in diatomic
 radicals with a $^2\Sigma_{1/2}$ ground state for such experiments and the
 HgF, HgH and BaF molecules were first studied semiempirically by Kozlov
 \cite{Kozlov:85}.  At the same time, the first two-step {\sl ab~initio}
 calculation of PNC effects in PbF initiated by Labzowsky was finished
 by Titov\etal\ \cite{Titov:85Dism,Kozlov:87}.  A few years later, Hinds started an
 experimental search for the electron EDM in the YbF molecule, on which the
 first result was obtained by his group in 2002 (Sussex, UK) \cite{Hudson:02},
 $d_e{=}(-0.2 \pm 3.2){\times}10^{-26} e{\cdot}{\rm cm}$.  Though that
 restriction is worse than the best current $d_e$ datum (from
 the Tl experiment, see above), nevertheless, it is limited by only counting
 statistics, as Hinds\etal\ pointed out in \cite{Hudson:02}.

 A new series of electron EDM experiments on YbF by Hinds' group (Imperial
 College, UK) are in progress and a new generation of electron EDM experiments
 using a vapor cell, on the metastable $a(1)$ state of PbO, is being prepared
 by
%tav2
   the
 group of DeMille (Yale, USA).  The unique suitability of PbO for searching
 for the elusive $d_e$ is demonstrated by the very high projected statistical
 sensitivity of the Yale experiment to the electron EDM.  In prospect, it
 allows one to detect $d_e$ of order of $10^{-29}\div10^{-31}\ e\cdot$cm
 \cite{DeMille:00}, two--four orders of magnitude lower than the current limit
 quoted above.  Some other candidates for the EDM experiments, in particular,
 HgH, HgF, TeO$^*$, and HI$^+$ are being discussed and an experiment on PbF is
 planned
%tav2 (Univ. of Oklahoma, USA).
 (Oklahoma Univ., USA).
%end
% One should mention also about promising PNC experiment on the Xe liquid
% \cite{Romalis:01}

%============================================================================
\section{Present status of the electron EDM search}
 \label{sStatus}
%============================================================================

% To date the search for the EDMs of elementary particles produced null
% results.  Nevertheless, there are serious reasons to search for the EDMs with
% the presently accessible (expected) level of the experimental sensitivity.
% The observation of a non-zero EDM points out
 As is  mentioned in the introduction, the observation of a non-zero EDM would point
 out the presence of so called ``new physics'' (see \cite{Okun:82, Erler:04}
 and references) beyond the Standard Model
%tav: to save the order of references:
% \cite{Glashow:61, Weinberg:67, Salam:68, Weinberg:72}
  \cite{Glashow:61, Weinberg:67, Salam:68, Weinberg:72, Kobayashi:73}
%end
 or CP violation in the QCD sector of SM, $SU(3)_C$.
 The discovery of a lepton EDM (electron EDM in our case) would have an
 advantage as compared to the cases of neutron or proton EDMs because the
%tav
% former need not be further interpreted at the quark-gluon level
% to be useful for SM extensions of \textbf{\textit{six-quark
% ``Cabbibo-Kobayashi-Maskawa'' model \cite{Kobayashi:73}}}.\marginpar{*
% I don't agree with this}
% to be useful for SM extensions.
  latter are not considered as elementary particles within the SM and its
  extensions.
%end

 In \Tref{teEDMsta} some estimates for the electron EDM predicted by different
 theoretical models are given (e.g., see \cite{Commins:99} for more details).
 One can see from the table that the most conservative estimate is given by the
 Standard Model.  This is explained by severe cancellations and suppressions of
 the contributions producing the electron EDM within the SM.  In turn, the
 ``new physics'' (extensions of the Standard Model:  supersymmetry (SUSY)
 \cite{Kazakov:00, Mohapatra:03, Erler:04} multi-Higgs \cite{Barr:92a,
 Barr:93b, Ginzburg:04}, left-right symmetry \cite{Pati:74, Barr:93b,
 Mohapatra:03}, lepton flavor-changing \cite{Liu:94, Masina:04}
 etc.) is very sensitive to the EDMs of elementary particles.
 This is especially true for the minimal (``naive'') SUSY model, which predicts
 an electron EDM already at the level of $10^{-25} e{\cdot}{\rm cm}$.
 However, the best experimental estimate on the electron EDM,
 $1.6{\times}10^{-27} e{\cdot}{\rm cm}$, obtained in the experiment on the
 Tl atom \cite{Regan:02}, is almost two orders of magnitude smaller.
 More sophisticated SUSY models (which are extremely popular among theorists
 because they allow one to overcome serious theoretical drawbacks of SM,
 explain the ``gauge hierarchy problem'', solve the problem of dark matter in
 astrophysics etc.) still predict the electron EDM at the level of $10^{-27}
 e{\cdot}{\rm cm}$ or somewhat smaller.  Since the Tl experiment is finished
 now, an intriguing expectation is connected with the ongoing experiment on
 the $a(1)$ state of the PbO molecule, which is expected to be sensitive to
 the electron EDM at least two orders of magnitude smaller.
% up to the level of $10^{-29}-10^{-31} e{\cdot}{\rm cm}$.
 Thus, the most popular extensions of SM can be severely examined by this
 experiment, i.e.\ even the result compatible with zero will dramatically
 influence their status.

%##########################################################################
%\renewcommand{\baselinestretch}{1}
%\clearpage
%\begin{table}[tb]
\begin{table*}
%\begin{table}
\caption{Prediction for the electron EDM, $|d_e|$, in popular theoretical
         models}
 \label{teEDMsta}

%\bigskip
\begin{center}
\begin{tabular}{lcc}
\hline
\hline
 \vspace{-2mm}\\
  {\bf Model}   &~~~~~~~&   $ |d_e|$ (in $e{\cdot}{\rm cm}$) \\
 \vspace{-2mm}\\
\hline
\hline
 \vspace{-3mm}\\
 Standard Model                        &&      $< 10^{-38}$          \\
 \vspace{-3mm}\\                     
\hline                                
 \vspace{-3mm}\\                     
 Left-right symmetric                  &&      $  10^{-28}-10^{-26}$ \\
 \vspace{-3mm}\\                     
\hline                                
 \vspace{-3mm}\\                     
 Lepton flavor-changing                &&      $  10^{-29}-10^{-26}$ \\
 \vspace{-3mm}\\                     
\hline                                
 \vspace{-3mm}\\                     
 Multi-Higgs                           &&      $  10^{-28}-10^{-27}$ \\
 \vspace{-3mm}\\                     
\hline                                
% \vspace{-3mm}\\                     
% Technicolor \cite{Lane:00}            &&   $\sim 10^{-29}$          \\
% \vspace{-3mm}\\
%\hline                        
 \vspace{-3mm}\\
 Supersymmetric                        &&   $\le  10^{-25}$          \\
 \vspace{-3mm}\\
\hline
\hline
 \vspace{-3mm}\\
 Experimental limit\cite{Regan:02} && $< 1.6{\times}10^{-27}$ \\
% ???\footnotemark[1]
 \vspace{-3mm}\\
\hline
\hline
\end{tabular}
\end{center}
%\medskip

%%\noindent
%\hspace{-3mm}
%\footnotetext[1]{}
%%\medskip
%\hspace{-3mm}
%\footnotetext[2]{}

%\end{table}
\end{table*}
%##########################################################################

%============================================================================
\section{General remarks on experimental search for EDMs in atoms and
         molecules}
 \label{sPTexp}
%============================================================================

 The experiments to search for EDMs in atoms and molecules are carried out
 using different approaches \cite{Khriplovich:97,Commins:99}.  The experimental
 technique depends on the properties of the atoms and molecules used in such an
 experiment.  These properties influence the atomic and molecular sources,
 resonance region and detector.  For example, for diatomic radicals like YbF or
 PbF the experiments on molecular beams are most reasonable,
%tav
 while for molecules with closed electronic shells in the ground state like PbO
 the EDM measurements can be carried out in optical cells.
%end tav

%\textbf{\textit ... }\marginpar{* You can change this}

 Nevertheless, the statistical sensitivity of the experiments to the electron
 or proton EDM usually depends on some parameters common for all such
 EDM experiments.  The easiest way to see this is to apply the Heisenberg
 uncertainty principle to evaluate the sensitivity of the EDM measurement.
 Suppose that the EDM of a molecule is measured in some electric field,
 $\vec{E}$.  (Do not confuse the {\it EDM of a polar molecule} with the large
 conventional {\it dipole moment} of the molecule, which averages to
 zero in the absence of external electric field in the laboratory coordinate
 system.  In contrast to the latter, the (vanishingly small) molecular EDM can exist
 only due to P,T-odd interactions; it is {\it permanent} and its direction
 depends on the sign of the projection of the total electronic momentum on the
 molecular axis. See \cite{Commins:99} for more details.) Thus the energy of
 interaction of the molecular EDM,
%tav: here and below $|\vec{d}|$ is replaced by $d$ etc.:
% $\vec{d}$,
  $\vec{d}={\it d}\vec{\sigma}$ (where $\vec{\sigma}$ is a unit vector
 along the total angular momentum of the molecule),
%end
 with the electric field is
 $\vec{d} \cdot \vec{E}$ (linear Stark effect) and the energy difference between
 the levels with opposite directions of the total angular momentum (leading to the
 contributions of different signs) is $2\vec{d} \cdot \vec{E}$.  If a
 measurement is carried out by detecting the energy shift during a time $T$,
 the uncertainty in the
%tav
% $|\vec{d}|$ determination is
  $d$ determination is
% $\delta |\vec{d}|= \hbar /(2 T \vec{E} \cdot \vec{\sigma})$, where $\vec{\sigma}$
% is a unit vector along the total angular momentum of the molecule.
%%$\delta {\it d}= \hbar /(2 T |\vec{E} \cdot \vec{\sigma}|)$.
  $\delta {\it d}= \hbar /(2 T \vec{E} \cdot \vec{\sigma})$.
%end
  For such measurement on $N$ uncorrelated molecules one, obviously, has
$$
%tav  \delta |\vec{d}| = \hbar /(2 T \sqrt{N} \vec{E} \cdot \vec{\sigma}) =
%%    \delta {\it d} = \hbar /(2 T \sqrt{N} |\vec{E} \cdot \vec{\sigma}|) =
      \delta {\it d} = \hbar /(2 T \sqrt{N} \vec{E} \cdot \vec{\sigma}) =
%end
      \hbar/(2 T E_\sigma \sqrt{\tau dN/dt})\ ,
$$
%
%tav ??? <correct later: d / |d| and corresponding values>
  where $dN/dt$ is the counting rate, $E_\sigma=\vec{E} \cdot \vec{\sigma}$, and
% where $dN/dt$ is the counting rate, $E_\sigma=|\vec{E} \cdot \vec{\sigma}|$, and
%end
 $\tau$ is the total measurement time (usually $\tau \gg T$ and $T$ is limited
 by the {\it coherence time} of the considered system).  Up to now we deal with
%tav
% the molecular EDM  $\vec{d}$.  Let us write $|\vec{d}|=G|\vec{d_e}|$, where
% $\vec{d_e}$ is the {\it electron EDM} (the same is valid, of course, for the
  the molecular EDM  $\vec{d}$.  Let us write \mbox{${\it d}=G{\it d}_e$}, where
  ${\it d}_e$ is the
% absolute
  value of the {\it electron EDM} (the same is valid, of course, for the
%end
 proton EDM) and $G$ is the proportionality coefficient (usually called the
 {\it enhancement factor}). Thus, the final expression for
%tav $\delta |\vec{d_e}|$
     $\delta {\it d}_e$
%end 
 is
\begin{equation}
%tav  \delta |\vec{d_e}| = \frac{\hbar}{2 T G E_\sigma \sqrt{\tau
  \delta {{\it d}_e} = \frac{\hbar}{2 T G E_\sigma \sqrt{\tau
%end
         dN/dt}}=\frac{\hbar}{2 T W \sqrt{\tau dN/dt}},
 \label{delta_d}
\end{equation}
 where the value $W = G E_\sigma$ is the {\it effective electric field} in the
 molecule, which can be interpreted as the field that should be applied along
 the EDM of a free electron
% (in parallel and antiparallel)
%tav
% to give the energy shift $2W|d_e| \equiv 2E_\sigma |d|$.
  to give the energy shift $2W {\it d}_e \equiv 2E_\sigma {\it d}$.
%end

 From expression \eref{delta_d}, the basic conditions which should be
 met in any prospective EDM experiment can be derived:
\begin{enumerate}
\item
   The counting rate ($dN/dt$) should be made as high as possible.  From this
   point of view the experiments on vapor cells, like that planned for PbO,
   have a clear advantage as compared to beam experiments because molecular
   vapor density can be usually made much higher than molecular beam density.
   Thus, in the experiment on the PbO cell the counting rate is estimated to be of
   order $10^{11}$--$10^{15}$~Hz \cite{DeMille:00}, while in the experiment on the YbF
   molecular beam the counting rate was of order 10$^3$---10$^4$~Hz
   \cite{Hudson:02}.
\item
   It is crucial to attain high coherence time $T$. In the beam experiments
   that time is just the time of flight through the region with the electric
   field. For a gas-dynamic molecular source the typical time of flight is
%tav corrected {\rm ms} etc.:
   $1-10$~ms. On the other hand, for the PbO experiment in vapor cell $T$ is
   close to the lifetime of the excited (metastable) state $a(1)$,
   $T \approx 0.1$~ms.
%end
   So, the beam experiments have advantage in the coherence time.
\item
   It is also critical to have a high value of the effective electric field $W$,
   acting on the electron. The only way to know that parameter is to perform
   relativistic calculations.  It is notable that the first semiempirical
   estimates of this kind were performed by Sandars in \cite{Sandars:65,
   Sandars:67} for Cs
   and TlF, correspondingly.  In these papers the importance of accounting for
   relativistic effects and using heavy atoms and heavy-atom molecules in EDM
   experiments was first understood.
\end{enumerate}

 The expected energy difference, $2\vec{d}\cdot\vec{E}$ is extremely small even
 for completely polarized heavy-atom molecules. Thus, in practice, the EDM
 experiment is usually carried out in parallel and antiparallel electric and
 magnetic ($\vec{B}$) fields. Interaction energy of the molecular magnetic
 moment, $\vec{\mu}$, with the magnetic field is much higher than that of the
 EDM with the electric field and the energy differences are
$$
   2\vec{\mu}\cdot\vec{B} + 2\vec{d}\cdot\vec{E}
$$ 
 and 
$$
   2\vec{\mu}\cdot\vec{B} - 2\vec{d}\cdot\vec{E}
$$
 for parallel and antiparallel fields, respectively (in practice, the
 atomic or molecular spin precession is usually studied instead of direct
 measurement of the energy shift, see \cite{Khriplovich:97}).  When the
 electric field is reversed, the energy shift,
%tav
% $4\vec{d}\cdot\vec{E} = 4|\vec{d_e}|\cdot W$,
  $4\vec{d}\cdot\vec{E} = 4{\it d}_e W$,
%end
 points to the existence of the permanent molecular EDM.  The same measurement
 technique is applicable to studying other P,T-odd interactions in atoms and
 molecules.

% The molecular
 The electronic structure
 parameters describing the P,T-odd interactions of electrons (sections
 \ref{sPbFHgF}, \ref{sYbFBaF}, and \ref{sPbO}) and nucleons (section
 \ref{sTlF}) including the interactions with their EDMs should be {\it
 reliably} calculated for interpretation of the experimental data.  Moreover,
 {\sl ab~initio} calculations of some molecular properties are usually required
 even for the stage of preparation of the experimental setup.  Thus,
 electronic structure calculations suppose
%tav2
   a
 high level of accounting for both
 correlations and relativistic effects (see below).  Modern methods of
 relativistic \abinitio calculations (including very recently developed
 approaches) allow one to achieve the required high accuracy.  These approaches
 will be outlined and discussed in the following sections.

%============================================================================
\section{Heavy-atom molecules: Computational strategies.}
 \label{sHAMcalc}
%============================================================================

 The most straightforward method for electronic structure calculation of
%tav molecules containing heavy atoms
 heavy-atom molecules
%end
 is solution of the eigenvalue problem using
 the Dirac-Coulomb (DC) or Dirac-Coulomb-Breit (DCB) Hamiltonians
 \cite{\DCBrev} when some approximation for the four-component wave function is
 chosen.

%tav: <SCF calculation is performed on a very large (atomic) basis set and only
%      relatively small number of the SCF solutions, ``moleculars spinors'',
%      are used later in correlation calculations, after transformation of
%      two-electron integral from atomic basis to the molecular one:>
%
% \marginpar{unclear}\textbf{\textit{However, even applying the four-component
% single configuration (SCF) approximation, Dirac-Fock (DF) or Dirac-Fock-Breit
% (DFB), to heavy-atom molecules followed by transformation of two-electron
% integrals is not always an easy task already because large sets of primitive
% atomic basis functions can be required for such all-electron four-component
% calculations.}}
  However, even applying the four-component single configuration (SCF)
  approximation, Dirac-Fock (DF) or Dirac-Fock-Breit (DFB), to calculation of
  heavy-atom molecules
  (followed by transformation of two-electron integrals to
  the basis of molecular spinors is not always an easy task)
%tav1:  already
  because a
  very large set of primitive atomic basis functions can be required for such
  all-electron four-component SCF calculations (see \cite{Mosyagin:05a}).
%end
 Starting from the Pauli approximation and Foldy--Wouthuysen transformation,
 many different two-component approaches were developed in which only large
 components are treated explicitly (e.g., see \cite{\TwoCompAppr} and
 references).  In addition, the approaches with perturbative treatment of
 relativistic effects \cite{Kutzelnigg:90} have been developed in which a
 nonrelativistic wavefunction is used as reference.  During the last few years,
 good progress was also attained in four-component techniques \cite{Dyall:02a,
 Visscher:02aa, Grant:04A, Schwerdtfeger:02bb} which allowed one to reduce
 efforts in calculation and transformation of two-electron matrix elements with
 small components of four-component molecular spinors.  These developments are
 applied, in particular, in the {\sc dirac} \cite{DIRAC} and {\sc bertha}
 \cite{Quiney:99,BERTHA} molecular programs.  Thus, accurate DC(B) calculations
 of relatively simple heavy-atom molecules can be performed on modern computers
 now.

 The greatest computational savings are achieved when the two-component
 relativistic effective core potential (RECP) approximation suggested
 originally by Lee\etal\ \cite{Lee:77} is used (e.g., see reviews
 \cite{\RECPrev}).  There are several reasons for using RECPs (including model
 potentials and pseudopotentials) in calculations of complicated heavy-atom
 molecules, clusters and solids. The RECP approaches allow one to exclude the
 large number of chemically inactive electrons from molecular calculations and
 to treat explicitly only valence and outermost core electrons from the
 beginning.  Then, the oscillations of the valence spinors are usually smoothed
 in heavy-atom cores simultaneously with excluding small components from the
 explicit treatment. As a result the number of primitive basis functions can be
 reduced
%tav2 dramatically
      dramatically;
 this is especially important for calculation and transformation of
 two-electron integrals when studying many-atomic systems and compounds of very
 heavy elements including actinides and superheavies.  The RECP method is based
 on a well-developed nonrelativistic technique of calculations; however, an
 effective spin-orbit interaction and other scalar-relativistic effects are
 taken into account usually by means of radially-local~\cite{\SemiLocRECP},
 separable \cite{\SeparablePP} or Huzinaga-type~\cite{\HuzinagaPP} operators.

 Correlation molecular calculations with RECPs are naturally performed in
 the basis of spin-orbitals (and not of spinors as is in all-electron
 four-component calculations) even for the cases when quantum electrodynamics
 (two-electron Breit etc.) effects are taken into account
 \cite{Petrov:04b, Mosyagin:05a}.
 Note, however, that the DCB technique with the separated spin-free and
 spin-dependent terms also has been developed \cite{Dyall:94}, but it
 can be efficiently applied only in the cases when spin-dependent effects
% are not significant 
  can be neglected
 both for valence and for core shells.  In the RECP method, the
 interactions with the excluded inner core shells (spinors!) are described by
 spin-dependent potentials whereas the explicitly treated valence and outer
 core shells are usually described by spin-orbitals in molecular calculations.
 It means that some ``soft'' way of accounting for the core-valence
 orthogonality constraints is applied in the latter case
%tav3
%  (see \cite{Titov:99} for more details).
   \cite{Titov:99} (note, meantime, that the strict core-valence orthogonality
   can be retrieved after the RECP calculation by using the restoration
   procedures described below).
%end tav3
 Another merit of the RECP method is in its natural ability to account for
 correlations with the explicitly excluded inner core electrons
 \cite{Mosyagin:04a} (this direction is actively developed during last years).
 The use of the molecular spin-orbitals and the ``correlated'' RECPs allows one
 to reduce dramatically the expenses at the stage of correlation calculation
 of heavy-atom molecules.  These are important advantages when
%tav2
   a
 very high level of accounting for correlations is required even in studying
 diatomics (e.g., see calculations of PbO described in section \ref{sPbO}).
 Thus, many complications of the DC(B) molecular calculations are avoided when
 employing RECPs.

%tav 
% The radially-local RECP approaches for ``shape-consistent'' (or
% ``norm-conserving'') pseudoorbitals are most widely applied in
% \marginpar{unclear}\textbf{\textit{calculations of mole\-cu\-les with heavy
% elements though ``energy-adjusted/consistent'' pseudopotentials
% \cite{Schwerdtfeger:03} by Stuttgart team and Huzinaga-type ``{\sl ab~initio}
% model potentials'' \cite{Seijo:04} are also actively used.}}
  The ``shape-consistent'' (or ``norm-conserving'') RECP approaches are most
  widely employed in calculations of heavy-atom molecules though
  ``energy-adjusted/consistent'' pseudopotentials \cite{Schwerdtfeger:03} by
  Stuttgart
%end 
%tav3 corrected:
% ``energy-adjusted/consistent'' and ``{\sl ab~initio} model
%   potentials'' are developed by different groups.>
  team are also actively used as well as the
  Huzinaga-type ``{\sl ab~initio} model potentials'' \cite{Seijo:04}.
%end tav3
 In plane~wave calculations of many-atom systems and in molecular dynamics, the
 separable pseudopotentials \cite{\SeparablePP} are more popular now because
 they provide linear scaling of computational effort with the basis set
%tav
% size.
  size in contrast to the radially-local RECPs.
%end
 The nonrelativistic shape-consistent effective core potential was first
 proposed by Durand\ande Barthelat \cite{Durand:75} and then a modified scheme
 of the pseudoorbital construction was suggested by Christiansen\etal\
 \cite{Christiansen:79} and by Hamann\etal\ \cite{Hamann:79}.

 In a series of papers (see \cite{Titov:99, Titov:00a, Titov:02Dism, Petrov:04b,
 Mosyagin:04a} and references)
%tav2
% the
  a
 generalized RECP approach was developed that involves both radially-local,
 separable and Huzinaga-type potentials as its components in particular cases.
 It allows one to attain very high accuracy of calculation of valence
 properties and electronic densities in the valence region when treating
 outermost core shells in calculations explicitly (see section \ref{sRECPs} for
 more details).

 Nevertheless, calculation of such properties as spin-dependent electronic
 densities near nuclei, hyperfine constants, P,T-parity nonconservation
 effects, chemical shifts etc.\ with the help of the two-component
 pseudospinors smoothed in cores is impossible.
 We should notice, however, that the above core properties (and the majority of
 other properties of practical interest which are described by the operators
 heavily concentrated within inner cores or on nuclei) are
  mainly determined by electronic densities of the valence and outer core
  shells near to, or on, nuclei.  The valence shells can be open or easily
  perturbed by external fields, chemical bonding etc., whereas outer core
  shells are
  noticeably polarized (relaxed) in contrast to the inner core shells.
 Therefore, accurate calculation of electronic structure in the valence and
 outer core region is of primary interest for such properties.

 For evaluation of the matrix elements of the operators concentrated on (or
 close
 to) nuclei, proper shapes of the valence molecular four-component spinors must
 be restored in atomic core regions after performing the RECP calculation of
 that molecule.  In 1959, a nonrelativistic procedure of restoration of the
 orbitals from smoothed Phillips--Kleinman pseudoorbitals was proposed
 \cite{Phillips:59} based on the orthogonalization of the latter to the
 original atomic core orbitals.  In 1985, Pacios\ande Christiansen
 \cite{Pacios:85} suggested a modified orthogonalization scheme in the case of
 shape-consistent pseudospinors.  At the same time, a simple procedure of
%tav
% one-center restoration (i.e.\ \textbf{\textit{NOCR}}\marginpar{need to define}
% procedure, see below)
  ``nonvariational'' one-center restoration (NOCR, see below)
%end
 employing the idea of
 generation of equivalent basis sets in four-component Dirac-Fock and
 two-component RECP/SCF calculations was proposed and first applied in
 \cite{Titov:85Dism} to evaluation of the P,T-odd spin-rotational Hamiltonian
 parameters in the PbF molecule.
 In 1994, a similar procedure was used by Bl\"ochl inside the augmentation
 regions \cite{Blochl:94} in solids to construct the transformation operator
 between pseudoorbitals (``PS'') and original orbitals (``AE'') in his
 projector augmented-wave method.

 All the above restoration schemes
%tav can be
  are
%end
 called ``nonvariational'' as
 compared to the ``variational'' one-center restoration (VOCR, see below)
 procedure proposed in \cite{Titov:92A,Titov:96}.
 Proper behavior of the molecular orbitals (four-component spinors)
 in atomic cores of molecules can be restored in the scope
 of a variational procedure if the molecular pseudoorbitals (two-component
 pseudospinors) match correctly the original orbitals (large components of
 bispinors) in the valence region after the molecular RECP calculation.
%tav
% \textbf{\textit{This condition is rather correct when the shape-consistent
% RECP is involved to the molecular calculation with explicitly treated
% outermost core orbitals and, especially, when the
% \underline{GRECP}\marginpar{need to define} operator is used as demonstrated
% in \cite{Titov:99, Mosyagin:05a}.}}\marginpar{passage unclear}
  As is demonstrated in \cite{Titov:99, Mosyagin:05a}, this condition is
  rather correct when the shape-consistent RECP is involved to the molecular
  calculation with explicitly treated outermost core orbitals and, especially,
  when the generalized RECP operator is used since the above matching condition
  is implemented at their generation.
%end 

 At the restoration stage, a one-center expansion in the spherical harmonics
 with numerical radial parts is most appropriate both for orbitals (spinors)
 and for the description of ``external'' interactions with respect to the core
 regions of a considered molecule.  In the scope of the discussed two-step
 methods for the electronic structure calculation of a molecule, finite nucleus
 models and quantum electrodynamic terms including, in particular,
 two-electron Breit interaction may be taken into account
 without problems \cite{Petrov:04b}.

 One-center expansion was first applied to whole molecules by
 Desclaux\ande Pyykk\"o in relativistic and nonrelativistic {H}artree-{F}ock
 calculations for the series {CH$_{4}$ to PbH$_{4}$} \cite{Desclaux:74b} and
 then in the Dirac-Fock calculations of CuH, AgH and AuH \cite{Desclaux:76c}
 and other molecules \cite{Desclaux:02}.  A large bond length contraction due
 to the relativistic effects was estimated.  However, the accuracy of such
 calculations is limited in practice because the orbitals of the hydrogen atom
 are reexpanded on a heavy nucleus in the entire coordinate space.  It is notable that the RECP and one-center expansion approaches were considered earlier
 as alternatives to each other \cite{Pitzer:79,Pyykko:79}.

 The applicability of the discussed two-step algorithms for calculation of
 wavefunctions of molecules with heavy atoms is a consequence of the fact that
 the valence and core electrons may be considered as two subsystems,
 interaction between which is described mainly by some integrated properties of
 these subsystems.  The methods for consequent calculation of the valence and
 core parts of electronic structure of molecules give us a way to combine the
 relative simplicity and accessibility both of molecular RECP calculations in
 gaussian basis set, and of relativistic finite-difference one-center
 calculations inside a sphere with the atomic core radius.

 The first two-step calculations of the P,T-odd spin-rotational Hamiltonian
 parameters were performed for the PbF radical about 20 years ago
 \cite{Titov:85Dism, Kozlov:87}, with a semiempirical accounting for the
 spin-orbit interaction.  Before, only nonrelativistic SCF calculation of the
 TlF molecule using the relativistic scaling was carried out \cite{Hinds:80a,
 Coveney:83}; here the P,T-odd values were underestimated by almost a factor of
 three as compared to the later relativistic Dirac-Fock calculations.  The
 latter were first performed only in 1997 by Laerdahl\etal\ \cite{Laerdahl:97}
 and by Parpia \cite{Parpia:97}. The next two-step calculation, for PbF and HgF
 molecules \cite{Dmitriev:92}, was carried out with the spin-orbit RECP part
 taken into account using the method suggested in \cite{Titov:92}.

 Later we performed correlation GRECP/NOCR calculations of the core properties
 in YbF \cite{Titov:96b}, BaF \cite{Kozlov:97}, again in YbF \cite{Mosyagin:98}
 and in TlF \cite{Petrov:02}.  In 1998, first all-electron Dirac-Fock
 calculations of the YbF molecule were also performed by Quiney\etal\
 \cite{Quiney:98} and by Parpia \cite{Parpia:98}.  Recently we finished
 extensive two-step calculations of the P,T-odd properties and hyperfine
 structure of the excited states of the PbO molecule \cite{Isaev:04,Petrov:05a}.
  One more two-step calculation of the electron EDM enhancement effect was
  performed very recently for the molecular ion HI$^+$ \cite{Isaev:04b}. 

  We would like to emphasize here that the all-electron Dirac-Fock
  calculations on TlF and YbF are, in particular, important for checking the
  quality of the approximations made in the two-step method.  The comparison of
  appropriate results, Dirac-Fock vs.\ RECP/SCF/NOCR, is, therefore, performed
  in papers \cite{Mosyagin:98,Petrov:02} and discussed in the present paper.

 In this paper, the main features of the two-step method are presented and
 PNC calculations are discussed, both those without accounting for correlation
 effects (PbF and HgF) and those in which electron correlations are taken into
 account by a combined method of the second-order perturbation theory (PT2) and
 configuration interaction (CI), or ``PT2/CI'' \cite{Dzuba:96} (for BaF and
 YbF), by the relativistic coupled cluster (RCC) method \cite{Kaldor:97,
 Landau:01c} (for TlF, PbO, and HI$^+$),
 and by the spin-orbit direct-CI method
 \cite{\SODCIp} (for PbO).  In the {\sl ab initio} calculations discussed here, the
 best accuracy of any current method has been attained for the hyperfine constants and P,T-odd
 parameters regarding the molecules containing heavy atoms.

%============================================================================
\section{Two-step method of calculation of core properties}
 \label{s2step}
%============================================================================

 The two-step method consists of a two-component molecular RECP calculation at
 the first step, followed by restoration of the proper four-component wave
 function in atomic cores at the second step.  Though the method was developed
 originally for studying core properties in heavy-atom molecules, it can be
 efficiently applied to studying the properties described by the operators
 heavily concentrated in cores or on nuclei of light atoms in other
 computationally difficult cases, e.g., in many-atom molecules and solids.
 The details of these steps are described below.

%============================================================================
\paragraph{Generalized RECP}
 \label{sRECPs}
%============================================================================

 When core electrons of a heavy-atom molecule do not play an active role, the
 effective Hamiltonian with RECP can be presented in the form
\begin{equation}
   {\bf H}^{\rm Ef}\ =\ \sum_{i_v} [{\bf h}^{\rm Schr}(i_v) +
          {\bf U}^{\rm Ef}(i_v)] + \sum_{i_v > j_v} \frac{1}{r_{i_v j_v}}\ .
 \label{Heff2}
\end{equation}
 This Hamiltonian is written only for a valence subspace of electrons
 which are treated explicitly and denoted by indices $i_v$ and $j_v$.  In
 practice, this subspace is often extended by inclusion of some outermost core
 shells for better accuracy but we will consider them as the valence shells
 below if these outermost core and valence shells are not treated
 using different approximations.
 In~\Eref{Heff2}, ${\bf h}^{\rm Schr}$ is the one-electron Schr\"odinger
 Hamiltonian
\begin{equation}
     {\bf h}^{\rm Schr}\ = - \frac{1}{2} {\vec \nabla}^2 
     - \frac{Z_{ic}}{r}\ ,
 \label{Schr}
\end{equation}
 where {$Z_{ic}$} is the charge of the nucleus decreased by the number of inner
 core electrons. ${\bf U}^{\rm Ef}$ in \eref{Heff2} is an RECP (relativistic 
 pseudopotential) operator that
 is usually written in the  radially-local (semi-local) \cite{Ermler:88} or
 separable (e.g., see \cite{Theurich:01} and references) approximations
 when the valence pseudospinors are smoothed in the heavy-atom cores.
 Contrary to the four-component wave function used in Dirac-Coulomb(-Breit)
 calculations, the pseudo-wave function in the RECP case can be both two- and
 one-component.  The use of the effective Hamiltonian \eref{Heff2} instead of
 all-electron four-component Hamiltonians leads to the question about its
 accuracy.  It was shown both theoretically and in many calculations (see
 \cite{Titov:99,Titov:02Dism} and references) that the typical accuracy of the
 radially-local RECPs is within 1000--3000~\cm\ for transition energies between
 low-lying states though otherwise is sometime stated
  (see~\cite{Dolg:00aa,Titov:00bmin}).

%tav
% \textbf{\textit{In our two-step calculations the generalized RECP operator
% \cite{Titov:99,Titov:00a} is used
% that includes the radially-local, separable and Huzinaga-type
% relativistic pseudopotentials as its components and special cases.}}\marginpar{unclear}
  In our two-step calculations the generalized RECP operator
  \cite{Titov:99,Titov:00a} is used that includes the operators of
% relativistic
  radially-local shape-consistent RECP, separable pseudopotential and
  Huzinaga-type model potential as its components.
%? Besides they can be considered as some limiting cases of GRECPs.
%end
 Additionally, the GRECP operator can include terms of other types, known as
 ``self-consistent'' and two-electron ``term-splitting'' corrections
 \cite{Titov:95,Titov:99,Titov:00a}, which are important particularly for
 economical (but precise!) treatment of transition metals, lanthanides and
 actinides.  With these terms, the accuracy provided by GRECPs can be even
 higher than the accuracy of the frozen core approximation (employing the same
 number of explicitly treated electrons) because they can account for
 relaxation of explicitly excluded (inner core) electrons
 \cite{Titov:99,Titov:02Dism}.  The theoretical background of the GRECP concept
 is developed in a series of papers \cite{Titov:99, Titov:00a, Titov:02Dism,
 Petrov:04b, Mosyagin:05a, Mosyagin:04a}. In contrast to other RECP
 methods, GRECP employs the idea of separating the space around a heavy atom
 into three regions: inner core, outer core and valence, which are first
 treated employing different approximations for each.  It allows one to attain
 practically any desired accuracy, while requiring moderate computational
 efforts since the overall accuracy is limited in practice by possibilities of
 correlation methods.

 When innermost core shells must be treated explicitly, the
 four-component versions of the GRECP operator can be used, in principle,
 together with the all-electron relativistic Hamiltonians.  The GRECP can
 describe here some quantum electrodynamics effects (self-energy, vacuum
 polarization etc.) thus avoiding their direct treatment.
 One more remark is that the two-component GRECP operator can be applied even
 to very light atoms when
 smoothing the large components of the four-component spinors only in the
 vicinity of the nucleus just to account for relativistic effects (the GRECP
 for hydrogen provides accuracy of treatment of very small relativistic
 contributions within 5\%).

%============================================================================
%\subsection
\paragraph{Nonvariational One-Center Restoration}
 \label{sNOCR}
%============================================================================

\begin{figure}
\includegraphics[scale=0.9]{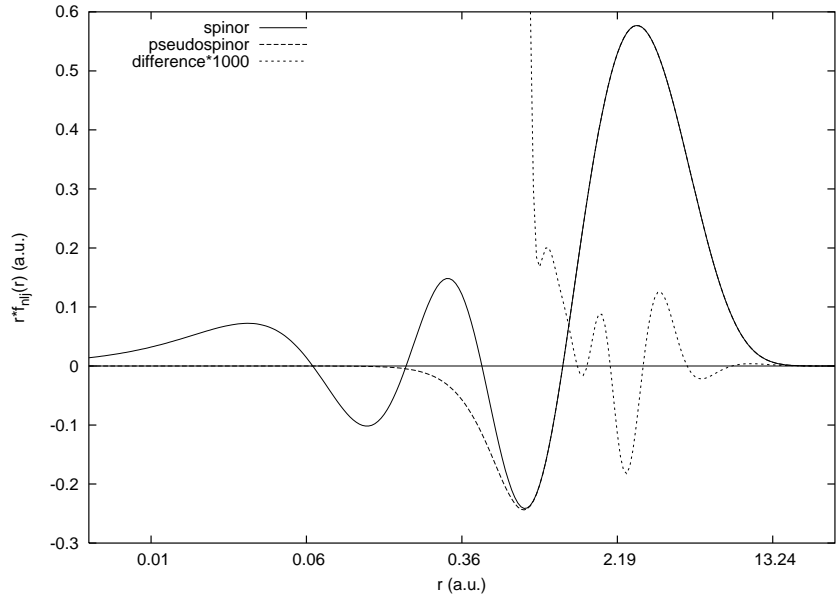}
\caption{\label{fTl_6p1}
 The radial parts of the large component of the $6p_{1/2}$ bispinor and the
 corresponding pseudospinor obtained in equivalent Dirac-Fock and 21-electron
 GRECP/SCF calculations for the state averaged over the relativistic
 $6s_{1/2}^2 6p_{1/2}^1$ configuration of thallium.  Their difference is
 multiplied by 1000.  The GRECP is generated for the state averaged over the
 nonrelativistic $6s^1 6p^1 6d^1$ configuration.
} \end{figure}

 The electronic densities evaluated from the two-component pseudo-wave function
 very accurately reproduce the corresponding all-electron four-component
 densities in the valence and outer core regions not only for the state used in
 the GRECP generation but also for other states which differ by excitations of
 valence electrons. This is illustrated in figure \ref{fTl_6p1} 
 (see also tables 8 and 9 in \cite{Mosyagin:05a}),
 where the radial parts of the large components of the thallium bispinor and
 the corresponding pseudospinor are compared for
 the state averaged over the relativistic $6s_{1/2}^2 6p_{1/2}^1$
 configuration, whereas the 21-electron GRECP is generated for the state
 averaged over the nonrelativistic $6s^1 6p^1 6d^1$ configuration.
 That is true also for the electronic densities obtained in the corresponding
 Dirac-Coulomb(-Breit) and GRECP calculations accounting for correlation.

 In the inner core region, the pseudospinors are smoothed, so that the
 electronic density with the pseudo-wave function is not correct.  When
 operators describing properties of interest are heavily concentrated near or
 on nuclei, their mean values are strongly affected by the wave function in the
 inner region.  The four-component molecular spinors must, therefore, be
 restored in the heavy-atom cores. 

 All molecular spinors $\phi _{p}$ can be restored as one-center expansions
 in the cores using the nonvariational one-center restoration (NOCR)
 scheme \cite{Titov:85Dism,\PToddMolCalc}
 that consists of the following steps:
\begin{itemize}
\item  Generation of {\it equivalent} basis sets of one-center four-component
       spinors
$
  \left(\begin{array}{c} f_{nlj}(r)\chi_{ljm}\\
              g_{nlj}(r)\chi_{2j{-}l,jm}\\
        \end{array}\right)
$
 and smoothed two-component pseudospinors \linebreak
$
    \tilde f_{nlj}(r)\chi _{ljm}
$
 in finite-difference all-electron Dirac-Fock(-Breit) and \linebreak GRECP/SCF
 calculations of {the same} configurations of a considered atom and its ions.
 The nucleus is usually modeled by a uniform charge distribution within a
 sphere.  The all-electron four-component {\sc hfdb} \cite{\cHFDB} and
 two-component
%tav: <it is defined in ref. \cHFJ [112]: HF in jj-coupling scheme. Besides,
%      the codes' names need not be defined:>
%
% {\sc  grecp/\underline{hfj}}\marginpar{define HFJ} \cite{\cHFJ}
  {\sc  grecp/hfj} \cite{\cHFJ}
%end
 codes are employed to generate two equivalent numerical basis sets used at the
 restoration.  These sets, describing mainly the atomic core region, are
 generated independently of the basis set used for the molecular GRECP
 calculations.

\item
 The molecular {\it pseudospinorbitals} are then expanded in the basis set of
 one-center two-component atomic {\it pseudospinors} (for $r{\le}R_{\rm nocr}$,
 where $R_{\rm nocr}$ is the radius of restoration that should be
 sufficiently large for calculating core properties with required accuracy),
\begin{equation}
    \tilde {\phi} _{p}({\bf x}) \approx
    \sum_{l=0}^{L_{max}}\sum_{j=|l-1/2|}^{j=|l+1/2|} \sum_{n,m}
    c_{nljm}^{p}\tilde f_{nlj}(r)\chi _{ljm}\ ,
 \label{expansion}
\end{equation}
 where ${\bf x}$ denotes spatial and spin variables.  Note that for linear
 molecules only one value of $m$ survives in the sum for each ${\phi} _{p}$.

\item
   Finally, the atomic two-component pseudospinors in the molecular basis
   are replaced by equivalent four-component spinors and the expansion
   coefficients from Eq.~(\ref{expansion}) are preserved:
\begin{equation} {\phi} _{p}({\bf x}) \approx
    \sum_{l=0}^{L_{\rm max}}\sum_{j=|l-1/2|}^{j=|l+1/2|} \sum_{n,m}
    c_{nljm}^{p}
     \left(
    \begin{array}{c}
    f_{nlj}(r)\chi _{ljm}\\
    g_{nlj}(r)\chi 
    _{2j-l,jm}
    \end{array}
    \right)\ .
 \label{restoration}
\end{equation}
\end{itemize}

 The molecular four-component spinors constructed this way are orthogonal to
 the inner core spinors of the atom,
% automatically 
 because the atomic basis functions
 used in Eq.~(\ref{restoration}) are generated with the inner core shells
 treated as frozen.

%============================================================================

%============================================================================
\paragraph{Variational one-center restoration}
 \label{sVOCR}
%============================================================================

 In the variational technique of one-center restoration (VOCR)
 \cite{Titov:92A,Titov:96}, the proper behavior of the four-component molecular
 spinors in the core regions of heavy atoms can be restored as an expansion in spherical harmonics inside the sphere with a restoration radius, $R_{\rm
 vocr}$, that should not be smaller than the matching radius, $R_c$, used at
 the RECP generation.  The outer parts of spinors are treated as frozen after
 the RECP calculation of a considered molecule.  This method enables one to
 combine the advantages of two well-developed approaches, molecular RECP
 calculation in a gaussian basis set and atomic-type one-center calculation in
 numerical basis functions,
%tav2
% by
  in
%end
 the most optimal way.  This technique is considered
 theoretically in \cite{Titov:96} and some results concerning the efficiency of
 the one-center reexpansion of orbitals on another atom can be found in
 \cite{Titov:02Dism}.

 The VOCR scheme can be used for constructing the core parts of the molecular
 spinors (orbitals) instead of using equivalent basis sets as in the NOCR
 technique.  A disadvantage of the NOCR scheme is that molecular
 pseudoorbitals are usually computed in a spin-averaged GRECP/SCF molecular
 calculation (i.e.\ without accounting for the effective spin-orbit
 interaction) and the reexpansion of molecular pseudospinorbitals on one-center
 atomic pseudospinors can be restricted
%tav2
% by
  in
%end
 accuracy, as
%tav2 it
 was noticed in the
 GRECP/RCC/NOCR calculations \cite{Petrov:02} of the TlF molecule (see below).
 With the restored molecular bispinors, the two-electron integrals on
 them can be easily calculated. Thus, the four-component transfomation from 
 the atomic basis that is a time-consuming stage of four-component 
 calculations of heavy-atom molecules can be avoided.
 Besides, the VOCR technique developed in \cite{Titov:96b} for molecular {\it
 pseudospinors} can be reformulated
%tav2
%  on the base
   for the case
%end
 of molecular {\it pseudospinorbitals} to reduce the complexity of the
 molecular GRECP calculation as is discussed in section \ref{sHAMcalc}.

 However, the most interesting direction in the development of the two-step
 method is the possibility to ``split'' the correlation structure calculation
 of a molecule into two sequential correlation calculations: first, in the
 valence region, where the {\it outer core and valence} electrons are
 explicitly involved in the GRECP calculation; and then, in the core region,
 when the {\it outer and inner core} space regions are only treated at the
 restoration stage.  Correlation effects in the valence and outer core regions
 (not only valence parts of molecular orbitals but also configuration
 coefficients) can be calculated, for example, by a combination of RCC and CI
 methods (see section \ref{sPbO}) with very high accuracy.  Then correlation
 effects in the inner and outer core regions
%tav added:
  (including the dipole-type relaxation of atomic inner core shells in a
  molecule)
%end
 can be taken into account using the Dirac-Coulomb(-Breit) Hamiltonian and the
 one-center expansion. 
%tav2
% Increasing the one-center restoration radius $R_{\rm vocr}$
  By increasing the one-center restoration radius $R_{\rm vocr}$\,,
%end
 one can take into account correlation effects in the intermediate
 region (outer core in our case) with the required accuracy.  Roughly speaking,
 the computational efforts for the correlation structure calculations in the
 core and valence regions are added when using the two-step approach, whereas
 in the conventional one-step scheme, they
%??? <better to rewrite:> 
 have multiplicative nature.

%============================================================================
\paragraph{Two-step calculation of molecular properties}
 \label{s2stProp}
%============================================================================

 To evaluate one-electron core properties (hyperfine structure, P,T-odd effects
 etc.) employing the above restoraton schemes it is sufficient to obtain the
 one-particle density matrix, $\{\widetilde{D_{pq}}\}$, after the molecular
 RECP calculation
%tav2
% on
  in
%end
 the basis of pseudospinors $\{\widetilde{\phi}_p\}$.  At the same time, the
 matrix elements $\{{W}_{pq}\}$ of a property operator $\bm{W}({\bf{x}})$
 should be calculated in the basis of equivalent four-component spinors
 $\{{\phi}_p\}$.  The mean value for this operator can be then evaluated as
\begin{equation} 
   \langle {\bm{W}} \rangle\ =\ \sum_{pq} \widetilde{D_{pq}} {W}_{pq}\ .
 \label{<W>} 
\end{equation} 
 Only the explicitly treated valence shells are used for evaluating a core
 property when applying \Eref{<W>} since the atomic frozen core (closed) shells
 do not usually contibute to the properties of practical interest.  However,
 correlations with the core electrons explicitly excluded from the RECP
 calculation can be also taken into account if the effective operator technique
 \cite{Lindgren:84} is applied to calculate the property matrix elements
 $\{{W}_{pq}^{\rm Ef}\}$
%tav2
% on
  in
%end
 the basis set of bispinors $\{{\phi}_p\}$.  Then,
 in expression \eref{<W>} one should only replace $\{{W}_{pq}\}$ by
 $\{{W}_{pq}^{\rm Ef}\}$.  Alternatively, the correlations with the inner 
 core electrons can be, in principle, considered within the variational 
 restoration scheme for electronic structure in the heavy atom cores.
 Strictly speaking, the use of the effective operators is correct when the
 molecular calculation is carried out with the ``correlated'' GRECP (see
 \cite{Mosyagin:04a}), in which the same correlations with the excluded core
 electrons are taken into account at the GRECP generation as they are in
 constructing $\{{W}_{pq}^{\rm Ef}\}$.  Nevertheless, the use of the (G)RECP
 that does not account for the core correlations at the molecular calculation
 stage can be justified if these correlations do not influence dramatically the electronic
 structure in the valence region.  The latter approximation was applied in
 the calculations of YbF and BaF molecules described in the following section.

\vspace{3mm}
 When contributions to $\langle {\bm{W}} \rangle$ are important both in the
 core and valence regions, the scheme for evaluating the mean value of
 $\bm{W}(\bf{x})$ can be as follows:
\begin{itemize}
\item 
 calculation of matrix elements with the molecular pseudospinorbitals (which
 are usually linear
%tav2
% combination
  combinations
%end
 of atomic gaussians) over the entire space region
 using conventional codes for molecular RECP calculations
%tav2
% (although,
  (although
%end
 the operator $\bm{W}$ can be presented in different forms in calculations with
 the RECP and Dirac-Coulomb(-Breit) Hamiltonians),
\begin{equation}
        \widetilde{\langle \bm{W} \rangle}\ =\ 
        \sum_{pq} \widetilde{D_{pq}}
                \int\limits_{r< \infty}
                \widetilde{\phi}_p({\bf{x}})\
                \bm{W}({\bf{x}})\
                \widetilde{\phi}_q({\bf{x}})\ d{\bf{x}}\ ;
 \label{12.1}
\end{equation} 
\item
 calculation of the inner part of the matrix element of the operator with the
 same molecular pseudospinorbitals using the one-center expansion for
 $\{\widetilde{\phi}_p\}$ ($R_{\rm ocr}$ stands for $R_{\rm nocr}$ or 
 $R_{\rm vocr}$ below, $R_{\rm ocr} \ge R_c$):
\begin{equation}
        {\widetilde{\langle \bm{W} \rangle}}^<\ =\ 
        \sum_{pq} \widetilde{D_{pq}}
                \int\limits_{ r < R_{\rm ocr}}
                \widetilde{\phi}_p({\bf{x}})\
                {\bm{W}}({\bf{x}})\
                \widetilde{\phi}_q({\bf{x}})\ d{\bf{x}}\ ;
 \label{12.2}
\end{equation} 
\item
 calculation of the inner part of the matrix element of the operator with the
 restored molecular four-component spinors using the one-center expansion for
 $\{{\phi}_p\}$:
\begin{equation}
        {\langle \bm{W} \rangle}^<\ =\ 
        \sum_{pq} \widetilde{D_{pq}}
                \int\limits_{r < R_{\rm ocr}}
                \phi^<_p (\bf{x})\
                {\bm{W}} (\bf{x})\
                \phi^<_q (\bf{x}) \ d\bf{x}\ .
 \label{12.3}
\end{equation}
\end{itemize} 
 The matrix element $\langle \bm{W} \rangle$ of the 
 $\bm{W}(\bf{x})$ operator is evaluated as
\begin{equation}
        {\langle \bm{W} \rangle}\ =\ \widetilde{\langle \bm{W}
                 \rangle}\ -\
                {\widetilde{\langle \bm{W} \rangle}}^<\
                +\ {\langle \bm{W} \rangle}^<\ .\
 \label{12.4}
\end{equation}
 Obviously, the one-center basis functions can be numerical (finite-difference)
 for better flexibility.

 The mean values of the majority of operators for the valence properties can be
 calculated with good accuracy without accounting for the inner parts of the
 matrix elements \eref{12.2} and \eref{12.3}.  As discussed above, for
 calculating the mean values of the operators heavily concentrated on or near
 nuclei it is sufficient to account only for the inner parts \eref{12.3} of the
 matrix elements of $\bm{W}(\bf{x})$ on the restored functions
 $\phi^<_p({\bf{x}})$, whereas the other contributions, \eref{12.1} and
 \eref{12.2}, can be neglected.

 Calculation of properties using the finite-field method
 \cite{Kunik:71,Monkhorst:77} and \Eref{<W>} within the coupled-cluster
 approach is described in section \ref{sTlF}.

%============================================================================
\section{Calculations of PbF and HgF}
 \label{sPbFHgF}
%============================================================================

 The ground states of the diatomic radicals PbF and HgF are $^2\Pi_{1/2}$ and
 $^2\Sigma_{1/2}$, respectively.  Here the superscript denotes spin multiplicity,
 $\Pi$ and $\Sigma$ are the projections of the electron orbital angular momentum on
 the molecular axis and the subscript is the projection of the total electron
 angular momentum on the molecular axis directed from the heavy atom to
 fluorine.  It is convenient to describe the spin-rotational spectrum of the
 ground electronic state in terms of the effective spin-rotational Hamiltonian
 $\opHeff^{sr}$, following \cite{Dmitriev:92,Kozlov:95}:
\begin{eqnarray}
 \label{Heff_PbFHgF}
\begin{array}{rl}
  \opHeff^{sr}& = B\vec{J}^2 + \Delta \vec{S}^{'}\cdot \vec{J} +
                      \vec{S}^{'}\cdot {\bf A} \cdot \vec{I}\\
    &+\ \mu_0 \vec{S}^{'}\cdot {\bf G} \cdot \vec{B} - D \vec{\lambda}\cdot \vec{E}\\
    &+\ W_{\rm A} k_{\rm A} \vec{\lambda}{\times}\vec{S}^{'}{\cdot}\vec{I}
     +\ (W_dd_e + W_{\rm S} k_{\rm S})\vec{S}^{'}{\cdot}\vec{\lambda}
\end{array}
\end{eqnarray}
 The first line in this expression describes the rotational structure with
 $\omega$- or spin-doubling and the hyperfine interaction of the effective
 electron spin $\vec{S}^{'}$ with the nuclear spin $\vec{I}$.  $B$ is the
 rotational constant, $\vec{J}$ is the electron-rotational angular momentum,
 $\Delta$ is the $\omega$-doubling constant.  The second line describes the
 interaction of the molecule with the external fields $\vec{B}$ and $\vec{E}$,
 ($\vec{\lambda}$ is the unit vector directed from the heavy nucleus to the
 light one).  The last line corresponds to the
 P-odd electromagnetic interaction of the electrons with the anapole moment of
 the nucleus described by the constant $k_{\rm A}$ \cite{Khriplovich:97},
 P,T-odd interaction of the electron EDM $d_e$ with the interamolecular field,
 and P,T-odd scalar interactions of the electrons with the heavy
 nucleus \cite{Dmitriev:92}.

 The parameter $\Delta$, tensors {\bf A} and {\bf G}, molecular dipole moment
 $D$ and the constants $W_i$ are expressed in terms of one-electron matrix
 elements; concrete expressions for the above parameters can be found in 
 \cite{Dmitriev:92}, and for $W_d$ and $A_\parallel$ they are also given in the
 next sections.  The results of the calculations are presented 
 in \Tref{tPbFHgF}.

%*************************************************************************
\begin{table*}
%\begin{table}
\caption{Parameters of the spin-rotational Hamiltonian for
         the ground states of $^{199}$HgF and $^{207}$PbF
	 ($\vec{I}$=1/2);  (a) experimental data \cite{Knight:81};
         (b) semiempirical results from \cite{Kozlov:85};
         (c) \abinitio calculations \cite{Dmitriev:92};
         (d),(e) \abinitio calculations \cite{Kozlov:87}
         with semiempirical accounting for the ``minimal'' and
         ``maximal'' spin-orbit mixing models, respectively,
         and $\Delta/2B$ from \cite{Huber:79}
%        , using $\Delta/2B=-0.3033$ \cite{Huber:79}, where
%        $B$ and $\Delta$ are rotational and $\omega$-doubling constants
         (the $W_i$ values in \cite{Kozlov:87} have wrong sign).}
 \label{tPbFHgF}

%***************************************************************************
%\bigskip
\begin{center}
\begin{tabular}{llrrrrrr}
\hline
\hline
\vspace{-4mm}\\
      &   &  $A_\parallel$  &$A_\perp$&$G_\parallel$ & $G_\perp$ & $W_{S}$ & $W_d$~~~~~~~~ \\
      &   & (MHz)           & (MHz)   &              &           & (kHz) &($10^{25}$\Hzecm)\\
\vspace{-4mm}\\
\hline
\hline
\vspace{-3mm}\\
 HgF  &(a)& 22621            & 21880 & 1.993      &  1.961  &      &   \\
      &(b)&                  &       &            &         & $-$191 & $-$4.8\\
      &(c)& 24150            & 23310 & 1.996      &  1.960  & $-$185 & $-$4.8\\
\hline
% PbF &(d)&                  &       &            &         &      &    \\
%     &(e)& 8690             & -7460 & 0.034      &  $-$0.269 & 51 & 1.0\\
%     &(f)& 9550             & -8240 & 0.114      &  $-$0.438 & 99 & 1.8\\
 PbF  &(d)& 8690             & -7460 & 0.034      &  $-$0.269 & 51 & 1.0\\
      &(e)& 9550             & -8240 & 0.114      &  $-$0.438 & 99 & 1.8\\
      &(c)& 10990            & -8990 & 0.040      &  $-$0.326 & 55 & 1.4\\

\hline
\hline
\end{tabular}
\end{center}

\end{table*}
%***************************************************************************

 In \cite{Dmitriev:92} the conclusion was made, that the accuracy in
 calculations of the parameters of the effective spin-rotational Hamiltonian
 is close to 20$\%$.  However, further \abinitio calculations showed the
 situation is more complicated.

 As was understood in calculations of YbF \cite{Titov:96b},
 a fortuitous cancellation of effects of different types took place in the
 above calculations.  Accounting for the
% spin-dependent 
 polarization of the $5s,5p$-shells leads to an enhancement of the
 contributions of the valence shells to the $A_\parallel$, $A_\perp$ and PNC
 constants on the level of 50\% of magnitude.  A contribution of comparable
 magnitude but of opposite sign takes place when the
 relaxation-correlation effects of the $5d$-shell are taken into account. This
 was confirmed in \cite{Mosyagin:XXa} when accounting for electron correlation
 both in the valence and core regions of HgF as compared to the YbF case.
% PT2/CI was used for HgF, RCC-SD for YbF and only SCF was used so far for HgH.
%<It maybe reasonable to add here PT2CI values on HgF>

%============================================================================
\section{Calculations of YbF and BaF}
 \label{sYbFBaF}
%============================================================================

 The results of two-step calculations for the YbF molecule
 (1996,1998)~\cite{Titov:96b,Mosyagin:98} and for the BaF molecule
 (1997)~\cite{Kozlov:97} are presented in \Tref{tBaFYbF}
%tav2
% \Tref{tBaFYbF}
  \Tref{tBaFYbF}, where they are
%end
 compared with other
 semiempirical and four-component results. For the isotropic hyperfine constant
 $A = (A_{\parallel}+2A_{\perp})/3$, the accuracy of our calculation is about
 3\%, as determined by comparison to the experimental datum.  The dipole constant $A_{\rm d} =
 (A_{\parallel}-A_{\perp})/3$ (which is much smaller in magnitude), though
 better than in all previous calculations known from the literature, is still
 underestimated by almost 23\%. After a semiemprical correction
 for a perturbation of $4f$-shell in the core of Yb due to the bond making,
 this error is reduced to 8\%. Our value for the effective electric field on
 the unpaired electron is
  $W = W_d|\vec{S}^{'}{\cdot}\vec{\lambda}|$ = 4.9~a.u.=
  $2.5 \times 10^{10}$ V\,cm$^{-1}$ (see section \ref{sPTexp} and
  \Eref{Heff_PbFHgF}).

 In \Tref{tBaFYbF} the values of the $W_d$ constant
\begin{equation}
        W_d d_{\rm e} =
        2 W d_{\rm e} =
        2 \langle ^2\Sigma_{1/2}|\sum_iH_{d}(i)| ^2\Sigma_{1/2} \rangle ,
 \label{e1a}
\end{equation}
 where $H_{d}$ describes interaction of the electron EDM
 $d_{\rm e}$ with the internal molecular
%tav2 ??? ${\bf E}^{\rm mol}$ is not well defined!
   electric
%end
 field ${\bf E}^{\rm mol}$:
\begin{eqnarray}
        &&H_{d} = 2 d_{\rm e}
        \left(\begin{array}{cc}
        0 & 0 \\
        0 & {\bf \sigma}
        \end{array} \right)
        \cdot {\bf E}^{\rm mol},
% \label{e1b}
\nonumber
\end{eqnarray}
 from various calculations are tabulated. These include the unrestricted
 Dirac-Fock calculation of Parpia (1998)~\cite{Parpia:98}, four-component
 calculations of Quiney\etal\ (1998)~\cite{Quiney:98} accounting for
 correlation, the most recent semiempirical calculation of Kozlov
 (1997)~\cite{Kozlov:94} and our latest GRECP/RAS\-SCF/EO calculation (EO
 stands for the effective operator technique in the framework of
%tav 
% \underline{PT2}\marginpar{define PT2},
  the second-order perturbation theory;
%end
 see section \ref{s2stProp} and paper \cite{Dzuba:96} for more details). All
 results are in
%tav2
% a
%end
 very close agreement now. It should be noted that the valence
 electron contribution to $W_d$ in~\cite{Parpia:98} differs by only 7.4\% from
 the corresponding RECP-based calculation of Titov\etal\
 (1996)~\cite{Titov:96b}.

 A similar increase in the values for the hyperfine constants and parameters of
 the P,T-odd interactions when the correlations with the core shells
%tav2
% (first of all, $5s,5p$)
  (primarily, $5s,5p$)
%end
 are taken into account is also observed for the BaF molecule
%tav
% \cite{Kozlov:97}
  \cite{Kozlov:97},
%end
 as one can see in~\Tref{tBaFYbF}.  Of course, the corrections from the
 $4f$-electron excitations are not required for this molecule.  The enhancement
 factor for the P,T-odd effects in BaF is three times smaller than in YbF
 mainly because of
%tav2
% smaller nuclear charge of Ba.  Therefore, barium fluoride can be considered
% as a less promising molecule to use for a search for PNC effects.
 the smaller nuclear charge of Ba.
%end

%\renewcommand{\baselinestretch}{1}
%##########################################################################
%\clearpage
%\begin{table}[tb]
\begin{table*}
%\begin{table}
\caption{The hyperfine structure constants
	 $A=(A_{\parallel}\,{+}\,2A_{\perp})/3$ (isotropic) and
         $A_d=(A_{\parallel}\,{-}\,A_{\perp})/3$ (dipole) and
         PNC parameters $W_i$ (i.e.\ $W_d$, $W_A$, and $W_S$)
         described in \Erefs{Heff_PbFHgF} and \eref{e1a}
         for the $^{171}$YbF and $^{137}$BaF molecules.}
 \label{tBaFYbF}

%\bigskip
%\begin{center}
\begin{tabular}{lcclcl}
\hline
\hline
\vspace{-4mm}\\
                &  $A$  &$A_{d}$&        \quad  $W_d$                & $W_A$ & $W_S$\\
                              & (MHz) & (MHz) & ($10^{25}$~\Hzecm) & (Hz) & (kHz)\\
\vspace{-4mm}\\
\hline
\hline
\multicolumn{6}{c}{\bf The $\phantom{\Bigl(}^{171}$YbF molecule} \\
\hline
\vspace{-3mm}\\
 Experiment~\cite{Knight:70}
                                        & 7617 & 102 &          &     &      \\
 Semiempirical~\cite{Kozlov:94}             &      &     & $-$1.5   & 730 & $-$48\\
 Semiempirical~\cite{Kozlov:97c}
                     (with $4f$-correction) &      &     & $-$1.26  &     & $-$43\\
\hline
 GRECP/SCF/NOCR \cite{Titov:96b}        & 4932 & ~59 & $-$0.91  & 484 & $-$33\\
 GRECP/RASSCF/NOCR \cite{Titov:96b}     & 4854 & ~60 & $-$0.91  & 486 & $-$33\\
\hline
 Restricted DF (Quiney, 1998) \cite{Quiney:98}\footnotemark[1]
%tav1
%                                       & 5918 & ~70 & $-$0.62  & 326 & $-$22\\
                                        & 5918 & ~35 & $-$0.62  & 326 & $-$22\\
 Rescaled\footnotemark[1]$^{,}$\footnotemark[2]
                      restricted DF     &      &     & $-$1.24  & 652 & $-$44\\
 Restricted DF + core polarization\footnotemark[1]
%                                       & 7865 & ~60 & $-$1.20  & 610 & $-$42\\
                                        & 7865 & ~60 & $-$1.20  & 620 & $-$42\\
%end tav1
\hline
 Unrestricted DF (Parpia, 1998) \cite{Parpia:98}\\
 ~~~ (unpaired valence electron)        &      &     & $-$0.962 &     &      \\
  Unrestricted DF \cite{Parpia:98}      &      &     & $-$1.203 &     & $-$44\footnotemark[1]\\
\hline
 GRECP/RASSCF/NOCR/EO \cite{Mosyagin:98}& 7842 & ~79 & $-$1.206 & 634 &      \\
 GRECP/RASSCF/NOCR/EO \cite{Mosyagin:98}\\
 ~~~         (with $4f$-correction) & 7839 & ~94 & $-$1.206 & 634 &  \\
\hline
\hline
\vspace{-4mm}\\
\multicolumn{6}{c}{\bf The $\phantom{\Bigl(}^{137}$BaF molecule} \\
\hline
\vspace{-3mm}\\
 Experiment~\cite{Knight:71}\footnotemark[3]
                                        & 2326 & 25 &          &     &       \\
 Semiempirical~\cite{Kozlov:85}
                                        &      &    & $-$0.41  & 240 &$-$13  \\
 Experiment~\cite{Ryzlewicz:82}\footnotemark[4]
                                        & 2418 & 17 &          &     &       \\
 Semiempirical~\cite{Kozlov:85}
                                        &      &    & $-$0.35  & 210 &$-$11  \\
\vspace{-4mm}\\
\hline
\vspace{-4mm}\\
 GRECP/SCF/NOCR \cite{Kozlov:97}        & 1457 & 11 & $-$0.230 & 111 &~$-$6.1\\
 GRECP/RASSCF/NOCR \cite{Kozlov:97}     & 1466 & 11 & $-$0.224 & 107 &~$-$5.9\\
 GRECP/SCF/NOCR/EO \cite{Kozlov:97}     & 2212 & 26 & $-$0.375 & 181 &       \\
 GRECP/RASSCF/NOCR/EO \cite{Kozlov:97}  & 2224 & 24 & $-$0.364 & 175 &       \\
%\vspace{-3mm}\\
%\vspace{-3mm}\\
\hline
\hline
\end{tabular}
%\end{center}
%\medskip
 \vspace{2mm}

\hspace{-3mm}
\noindent $^{\rm 1}$
%tav2                {Results
                     {The results
                      of Quiney\etal\ and Parpia
                      have been adjusted by a factor of
                      two to be consistent with the definition of $W_i$ used
                      here, see \Erefs{Heff_PbFHgF} and \eref{e1a}.}

\hspace{-3mm}
\noindent $^{\rm 2}$ {The $W_i$ values are rescaled from the ``restricted DF''
                      results employing the calculated and experimental $A$ and
                      $A_{d}$ values by the factor
 $\sqrt{(A^{\rm expt}{\cdot}A_d^{\rm expt})/(A^{\rm calc}{\cdot}A_d^{\rm calc})}$,
                      which are in good agreement with the
                      ``restricted DF\,+\, core polarization'' values.}

\hspace{-3mm}
\noindent $^{\rm 3}$ {The hyperfine constants are measured in an inert gas
                      matrix \cite{Knight:71} and the semiempirical
                      $W_i$ values \cite{Kozlov:85} are based on these data.}

\hspace{-3mm}
\noindent $^{\rm 4}$ {The hyperfine constants are measured in a molecular beam 
                     \cite{Ryzlewicz:82}.}
%\end{table}
\end{table*}

%============================================================================
\section{Calculation of~ $^{205}$TlF molecule.}
\label{sTlF}
%============================================================================

%=================================
\paragraph*{Effective Hamiltonian.}
 \label{sTlFHeff}
%=================================

 Here we consider the P,T-odd interaction of the $^{205}$Tl nucleus
%tav2
% having one unpaired proton
  (which has one unpaired proton) 
%end
 with the electromagnetic field of the electrons in the $^{205}$TlF molecule
 \cite{Petrov:02}.  This molecule is one of the best objects for
%tav2
% the proton EDM measurements. The effective interaction with the Tl nucleus
% EDM
  measurements of the proton EDM. The effective interaction with the EDM of the
  Tl nucleus  
%end
 in TlF is written in the form
\begin{equation}
     H^{\rm eff}=(d^V+d^M) \vec{I}/I \cdot \vec{\lambda}\ ,
 \label{interaction}
\end{equation}
%tav2 ", => ;" to separate definitions:
 where $\vec{I}$ is the Tl nuclear spin operator; $\vec{\lambda}$ is the unit
 vector along $z$ (from Tl to F); $d^V$ and $d^M$ are the {\it volume} and
 {\it magnetic} constants \cite{Schiff:63}:
\begin{equation}
   d^V=6SX=(-d_pR+Q)X\ ,
%\vspace{3mm}
 \label{dv}
\end{equation}
   where
 $S$ is the nuclear Schiff moment; $d_p$ is the proton EDM;
 $R$ and $Q$ are the factors determined by the
%***  Q is P,T-odd; R is P,T-even  ***
 {nuclear} structure of $^{205}$Tl (see \cite{Petrov:02});
\begin{equation}
   X=\frac{2\pi}{3} \left[
     \frac{\partial}{\partial z}\rho_{\psi}(\vec{r})
      \right] _{x,y,z=0}\ ;
  \label{X}
\end{equation}
 $\rho_{\psi}(\vec{r})$ is the electronic density calculated from the
 electronic wavefunction~$\psi$,
\begin{equation}
   d^M = 2 \sqrt{2}(d_p+d_N)
   \left(
   \frac{\mu}{Z}+ \frac{1}{2mc}
   \right)M\ ,
%\vspace{3mm}
 \label{dm}
\end{equation}
 where $d_N$ is the nuclear EDM arising due to P,T-odd
 forces between the nucleons;
 $\mu$, $m$ and $Z$ are the magnetic moment, mass and charge of the Tl
 nucleus; $c$ is the velocity of light;
\begin{equation}
   M = \frac{1}{\sqrt{2}}\langle\psi |\sum_i
   \left(\frac{\vec{\alpha}_i \times \vec{\bf l}_i}{r_i^3}\right)_{z}
   |\psi\rangle\ ;
 \label{M}
\end{equation}
 $\vec{\bf l_i}$ is the orbital momentum of $i$-th electron;
 $\vec{\alpha}_i$ are its Dirac matrices.
%end tav2
 Accounting for $H_{\rm eff}$ leads to
%tav2
% different
  a difference in the
 hyperfine splitting of TlF in
 parallel and antiparallel electric and magnetic fields. The level shift
  $h\nu = 4(d^V+d^M)\langle \vec{I}{\cdot}\vec{\lambda} \rangle$/I
 is measured experimentally (for the latest results
% on TlF
 see \cite{Cho:91}).

 The parameters $X$ of \Eref{X} and $M$ of \Eref{M} are determined by the
 electronic structure of the molecule. They were calculated in 1997 for the
 ground $0^+$ state of TlF by Parpia \cite{Parpia:97} and by
 Laerdahl, Saue, {Faegri~Jr.}, and Quiney \cite{Laerdahl:97}
 using the Dirac--Fock method with gaussian basis sets of large sizes (see
 Table \ref{result}).
 Below we refer to paper \cite{Quiney:98b} with the
 calculations presented in details and not to the brief communication
 \cite{Laerdahl:97} of the same authors.
 There was also a 
 preliminary calculation of electronic structure in TlF performed by
 Wilson\etal\ in \cite{Wilson:94}.  In paper \cite{Petrov:02} the first
 calculation of $^{205}$TlF that accounts for correlation effects was performed
 (see also \cite{Dzuba:02} where the limit on the Schiff moment of $^{205}$Tl
 was recalculated).

%===================
\paragraph*{Results.}
 \label{sTlFres}
%===================

 Calculations were carried out at two internuclear separations, the equilibrium
 $R_e=2.0844$ \AA\, as in Ref.\ \cite{Parpia:97}, and 2.1 \AA, for comparison
 with Ref.\ \cite{Quiney:98b}.
 The relativistic coupled cluster (RCC) method \cite{Kaldor:99,Kaldor:04b} with
 only single (RCC-S) or with single and double (RCC-SD) cluster amplitudes is
 used (for review of different coupled cluster approaches see also
 \cite{Paldus:99,Paldus:03} and references).  The RCC-S calculations with the
 spin-dependent GRECP operator take into account effects of the spin-orbit
 interaction at the level of the one-configurational SCF-type method. The
 RCC-SD calculations include, in addition, the most important electron
 correlation effects.

 The results obtained with the one-center expansion of the molecular spinors
 in the Tl core in either $s;p$, $s;p;d$ or $s;p;d;f$ partial waves are
 collected in Table \ref{result}.  The first point to notice is the difference
 between spin-averaged SCF values and RCC-S
%tav2
% values,
  values;
%end
 the latter include spin-orbit interaction effects. These effects increase $X$
 by 9\% and decrease $M$ by 21\%.  The RCC-S function can be written as a
 single determinant, and results may therefore be compared with DF values, even
 though the RCC-S function is not variational. The GRECP/RCC-S values of $M$
 indeed differ only by 1--3\% from the corresponding DF values
 \cite{Parpia:97,Quiney:98b} (see \Tref{result}).
 
 Much larger differences occur for the $X$ parameter.  There are also large
 differences between the two DF calculations for $X$, which cannot be explained
 by the small change in internuclear separation. The value of $X$ may be
 expected to be less stable than $M$ \cite{Quiney:98b}.  The conclusion in
 \cite{Petrov:02} is that the RCC-S value for $X$, which is higher than that of
 \cite{Parpia:97,Quiney:98b}, is more correct.  The electron correlation
 effects are calculated by the RCC-SD method at the molecular equilibrium
 internuclear distance $R_e$. A major correlation contribution is observed,
 decreasing $M$ by 17\% and $X$ by 22\%.

 The hyperfine structure constants of Tl $6p_{1/2}^1$ and Tl$^{2+}$ $6s^1$,
 which (like $X$ and $M$) depend on operators concentrated near the Tl
 nucleus, were also calculated. The errors in the DF values are 10--15\% 
 with respect to experiment and the RCC-SD results are within 1--4\% of 
 experiment. The improvement in $X$ and $M$ upon inclusion of correlation 
 is expected to be similar.

%\samepage
%***************************************************
%\begin{widetext}
% \squeezetable
% \begin{table*}[!]
\begin{table*}
\caption
 {Calculated $X$ \eref{X} and $M$ \eref{M} parameters (in a.u.) for the
 $^{205}$TlF ground state, compared with DF values
 \protect\cite{Parpia:97,Quiney:98b}. GRECP/RCC-S results include spin-orbit
 interaction, and GRECP/RCC-SD values also account for electron correlation.
}
\medskip
%\vspace{0.2cm}
% \begin{ruledtabular}
\begin{center}
% \begin{tabular}{ll|ddd|r|dd|r}
\begin{tabular}{ll|ccc|r|cc|r}
\hline
\hline
                          & & \multicolumn{4}{c|}{$R_e=2.0844$ \AA} &
                          \multicolumn{3}{c}{$R=2.1$ \AA} \\
\hline
 \multicolumn{2}{c|}{Expansion}
 & $s,p$ & $s,p,d$ & $s,p,d,f$ & $s,p$ & $s,p$ & $s,p,d,f$ & $s,p$ \\
\hline
 \multicolumn{2}{l|}{}
 & \multicolumn{3}{c|}{$M$} & $X$ & \multicolumn{2}{c|}{$M$} & $X$    \\
\hline
 \multicolumn{2}{l|}{SCF(spin-averaged)}
             & 19.67 &  17.56 &  17.51 &  8967   & 19.52 &  17.43 &  8897   \\
\hline
 \multicolumn{2}{l|}{GRECP/RCC-S}
             & 16.12 &        &  13.84 &  9813   & 16.02 &  13.82 &  9726   \\
\hline
 DF \cite{Parpia:97}&    &15.61 & &     &  7743  &     &     &  \\
\hline
 DF \cite{Quiney:98b}&  & & &   &  &&  13.64\footnotemark[1]& 8747 \\
\hline
 \multicolumn{2}{l|}{GRECP/RCC-SD}
             &       &        &  11.50 &  7635   &       &        &\\
\hline
\hline
\end{tabular}
\end{center}
% \end{ruledtabular}
 \vspace{2mm}

\hspace{-3mm}
 \noindent $^{\rm 1}\  M$
 is calculated in Ref.\ \cite{Quiney:98b} using two-center
 molecular spinors, corresponding to infinite $L_{max}$ in
 Eq.~(\ref{restoration}).

\label{result}
\end{table*}
%\end{widetext}

%============================================================================
\section{Calculations of~ $^{207}$PbO molecule.}
 \label{sPbO}
%============================================================================

 As is noted in section~\ref{sPThist}, experiments on the excited $a(1)$
 \cite{DeMille:00} or $B(1)$ \cite{Egorov:01} states of PbO having nonzero
 projection of total electronic momentum on the internuclear axis can be, in
 principle, sensitive enough to detect $d_e$ two or even four orders of
 magnitude lower than the current limit. Knowledge of the effective
 electric field, $W$, is required for extracting $d_e$ from the measurements
 (see section \ref{sPTexp}).  In papers \cite{Isaev:04,Petrov:05a}, $W$ for the
 $a(1)$ ($^3\Sigma^+$ $\sigma_1^2\sigma_2^2\sigma_3^2 \pi_1^3 \pi_2^1$) and
 $B(1)$ ($^3\Pi_1$ $\sigma_1^2\sigma_2^2\sigma_3^1 \pi_1^4 \pi_2^1$) states of
 the PbO molecule was calculated by the RCC-SD \cite{Kaldor:97,Landau:01c} and
 configuration interaction
%tav2
% CI
  (CI)
%end
 \cite{Buenker:74,\SODCIp} methods.  To provide an accuracy check for the
 calculation of the electronic structure near the Pb nucleus the hyperfine
 constant, $A_{\parallel}$, was also calculated.

%===================
\paragraph*{Results.}
 \label{sPbOres}
%===================

  CI calculations \cite{Petrov:05a} were performed at two internuclear
  distances: $R=3.8$ a.u. (as in RCC calculations), and $R=4.0$ a.u. (which is
  close to the equilibrium distances of the $a(1)$ and $B(1)$ states).  In the RCC calculations \cite{Isaev:04} the internuclear distance
  $R=3.8$ a.u.\ was used because of problems with convergence at larger
  distances.
 The calculated values with the one-center expansion of the molecular spinors
 in the Pb core in either $s;p$ or $s;p;d$ partial waves are collected in
 \Tref{res1}.

 Let us consider the $5s,5p,5d$ orbitals of lead and $1s$ orbital of oxygen
 as the outercore and the $\sigma_1$, $\sigma_2$, $\sigma_3$, $\pi_1$, $\pi_2$
 orbitals of PbO (consisting mainly of $6s,6p$ orbitals of Pb and $2s,2p$
 orbitals of O) as valence.  Although in the CI calculations we take into
 account only the correlation between valence electrons, the accuracy attained
 in the CI calculation of $A_{\parallel}$ is much better than in the RCC-SD
 calculation.  The main problem with the RCC calculation was that the 
 Fock-space RCC-SD version used there was not optimal in accounting for nondynamic
 correlations (see \cite{Isaev:00} for details of RCC-SD and CI calculations of
 the Pb atom).
% The crucial role of the nondynamic correlations became
% clear after corresponding SODCI calculations on PbO.
 Nevertheless, the potential of the RCC approach for electronic structure
 calculations is very high, especially in the framework of the
 intermediate Hamiltonian formulation \cite{Landau:01c,Kaldor:04b}.

 Next we estimate the contribution from correlations of valence electrons with
 outercore ones (which also account for correlations between outercore
% pairs of
 electrons) as the difference between the results of the corresponding 10- and
 30-electron GRECP/RCC calculations (see also \cite{Isaev:00} where this
 correction is applied to the Pb atom).  We designate such correlations in
 Table \ref{res1} as ``outercore correlations''. When taking into account
 outercore contributions at the point $R=4.0$ a.u.\ we used the results of the
 RCC calculation at the point $R=3.8$ a.u. Since these contributions are
 relatively small and because the correlations with the outercore electrons are
 more stable than correlations in the valence region when the internuclear
 distance is changed, this approximation seems reasonable.
 Finally, we have linearly extrapolated the results of the calculations to the
 experimental equilibrium distances, $R_e=4.06$ a.u.\ for $a(1)$
 \cite{Martin:88} and $R_e=3.91$ a.u.\ for $B(1)$ \cite{Huber:79}.  The final
 results are: $A_{\parallel} = -3826$~MHz, $W = -6.1{\cdot}10^{24}{\rm
 Hz}/(e\cdot {\rm cm})$ for $a(1)$ and $A_{\parallel} = 4887$~MHz, $W =
 -8.0{\cdot}10^{24}{\rm Hz}/(e\cdot {\rm cm})$ for $B(1)$.  The
 estimated error for the final $W$ value is 20\% for the $B(1)$ state.  For
 $a(1)$ the estimated error bounds put the actual $W$ value between 90\% and
 130\% of our final value (for details see \cite{Petrov:05a}).

%==========================================================================
\begin{table*}
\caption
{
 Calculated  parameters $A_{\parallel}$ (in MHz) and $W$
  (in $10^{24}{\rm Hz}/(e\cdot {\rm cm})$, see section \ref{sPTexp} and
  \Eref{e1a})
 for the $a(1)$ and $B(1)$ states of $^{207}$PbO at the
 internuclear distances 3.8 and 4.0 a.u.  The experimental value of
 $A_{\parallel}$ in $a(1)$ is $-4113$\,MHz \cite{Hunter:02}. The preliminary
 experimental value of $A_{\parallel}$ for $B(1)$ is $5000 \pm 200$\,MHz
 \cite{Kawall:04a}.
}
\medskip
% \begin{ruledtabular}
\begin{tabular}{lrrrrrrrrr}
\hline
\hline
State &\multicolumn{4}{c}{$a(1)$\ \ $\sigma_1^2\sigma_2^2\sigma_3^2 \pi_1^3
 \pi_2^1$ \ \  $^3\Sigma_1$}&
&\multicolumn{4}{c}{$B(1)$\ \ $\sigma_1^2\sigma_2^2\sigma_3^1 \pi_1^4
 \pi_2^1$\ \ $^3\Pi_1$}\\
 Parameters &\multicolumn{1}{c}{$A_{\parallel}$} &&\multicolumn{2}{c}{$W$}
& & \multicolumn{1}{c}{$A_{\parallel}$}&& \multicolumn{2}{c}{$W$}\\
% \cline{2-2}\cline{6-7}\cline{9-11}\cline{13-14}
 Expansion & s,p,d&& s,p& s,p,d& & s,p,d&& s,p& s,p,d \\

\hline
\multicolumn{10}{c}{ }\vspace{-2mm} \\
\multicolumn{10}{c}{\bf Internuclear distance $R=3.8$ a.u. } \\
\multicolumn{10}{c}{ }\vspace{-2mm} \\

%hline

 10e-RCC-SD \cite{Isaev:04}

           & -2635  && -2.93& -3.05 &  &3878&& -11.10 & -10.10 \\

 30e-RCC-SD \cite{Isaev:04}

           & -2698  &&      & -4.10 &      & 4081&& -9.10 &-9.70 \\

{\it outercore}: (30e-10e)-RCC-SD

           & \it -63&&     & \it -1.05&   & \it 203&&  &
\it 0.40  \\

\hline
10e-CI \cite{Petrov:05a}
            &     -3446 &&    &     -4.13 &       &     4582&&       &
-10.64 \\
\bf FINAL \cite{Petrov:05a}
            &            &&       &           &   &         &&  &        \\
(10e-CI + outercore)
            & \bf -3509 && & \bf -5.18  &      & \bf 4785 &&       & \bf -10.24 \\
\multicolumn{10}{c}{ }\vspace{-2mm} \\
\multicolumn{10}{c}{\bf Internuclear distance $R=4.0$ a.u. } \\
\multicolumn{10}{c}{ }\vspace{-2mm} \\

%\hline

 10e-CI \cite{Petrov:05a}
            &     -3689  &&       &     -4.81 &   &    4762 &&  &    -7.18 \\
 {\bf FINAL} \cite{Petrov:05a}
            &            &&       &           &   &         &&  &        \\
 (10e-CI + outercore)\footnotemark[1]
            & \bf -3752  &&       & \bf -5.86 &   &\bf 4965 &&  &\bf -6.78 \\
\hline
\hline
\end{tabular}
 \vspace{2mm}
% \end{ruledtabular}

\hspace{-3mm}
 \noindent $^{\rm 1}$\,{It is assumed that the outercore contribution at
 the internuclear distance
 $R{=}4.0$ a.u.\ is approximately the same as is at $R{=}3.8$ a.u.} 
\label{res1}
\end{table*}

%============================================================================
\section{Conclusions}
 \label{sConcl}
%============================================================================

 The P,T-parity nonconservation parameters and hyperfine constants have been
 calculated for the particular heavy-atom molecules which are of primary interest for
 modern experiments to search for PNC effects.  It is found that a high level
 of accounting for electron correlations is necessary for reliable
 calculation of these properties with
%tav2
  the
%end
 required accuracy.  The applied two-step
 (GRECP/NOCR) scheme of calculation of the properties described by the
 operators heavily concentrated in atomic cores and on nuclei has proved to be
 a very efficient way to take account of these correlations with moderate
 efforts.  The results of the two-step calculations for hyperfine constants
 differ by less than 10\% from the corresponding experimental data.  A
 comparable level of accuracy is expected for the P,T-odd spin-rotational
 Hamiltonian parameters, which can not be obtained experimentally.

 The precision of the discussed calculations is limited by the current
 possibilities of the correlation methods and codes and not by the GRECP and
 NOCR approximations, despite the fact that the GRECP/NOCR method allows one to reduce
 seriously the computational expenses of the correlation treatment as compared
 to conventional Dirac-Coulomb(-Breit) approaches when
 {\it (1)} using molecular spin-orbitals instead of spinors in (G)RECP
  calculation;
 {\it (2)} employing ``correlated'' GRECP versions \cite{Mosyagin:04a} to
  account for correlations with the core electrons explicitly excluded from
  (G)RECP calculations;
 {\it (3)} combining gaussian basis functions at the molecular (G)RECP
  calculation with numerical functions at the one-center restoration;
  and
 {\it (4)} splitting the correlation treatment of a molecule into two
  sequential steps, first in valence and then in core regions.

 In turn, the accuracy attained in the calculations presented above is
 sufficient for a reliable interpretation of the measurements in PNC
 experiments on these molecules.

%============================================================================
 
\vspace{4mm}
 \paragraph*{Acknowledgments.}

 We are very grateful to L.N.Labzowsky for initiating and supporting our
 activity in studying PNC properties in heavy-atom molecules over many years.
 We would like to thank our colleagues
%tav2 alphabethic ordering now:
% M.G.Kozlov, I.I.Tupitsyn, I.V.Abarenkov, U.Kaldor, E.Eliav, R.J.Buenker, and
% A.B.Alekseyev
 I.V.Abarenkov, A.B.Alekseyev, R.J.Buenker, E.Eliav, U.Kaldor, M.G.Kozlov,
 A.I.Panin, A.B.Tulub, and I.I.Tupitsyn
%end
 for many stimulating discussions and fruitful collaboration on the relevant
 research.

 The present work is supported by the U.S.\ CRDF grant RP2--2339--GA--02 and
 RFBR grant 03--03--32335.  A.P.\ is grateful to Ministry of education of
 Russian Federation (grant PD\,02--1.3--236) and to St.-Petersburg Committee of
 Science (grant PD\,03-1.3-60).  N.M.\ is also supported by the grants of
 Russian Science Support Foundation and the governor of Leningrad district.
%tav
% \marginpar{NSF}
  D.D.\ acknowledges additional support from NSF Grant No.\ PHY-0244927 and the
  David and Lucile Packard Foundation.
%end
%tav2
 A part of calculations of PbO was performed on computers of Boston University
 in the framework of the MARINER project.
%end

%============================================================================
%-\bibliographystyle{apsrev}            %%% short JPB-like (bad)
%-\bibliographystyle{apsrmp}
%-\bibliographystyle{apsr}
%-\bibliographystyle{abbrvnat}          %%% long
% \bibliographystyle{apalike}           %%% long JPB-like (good)
% \bibliographystyle{jmr}               %%% long

%-\bibliographystyle{jphysicsB}
%-\bibliographystyle{kluwer}
%-\bibliographystyle{amsalpha}
%-\bibliographystyle{alpha}
%-\bibliographystyle{agsm}              %%%     bad

% \bibliographystyle{tav1d2}            %%% long *** best so far ***
% \bibliographystyle{tav}               %%% long

% \bibliographystyle{unsrt}             %%% long
% \bibliographystyle{abbrv}             %%% long
% \bibliographystyle{plain}             %%% long
% \bibliographystyle{acm}               %%% long
% \bibliographystyle{siam}              %%% long []+alphab. 
% \bibliographystyle{ieeetr}            %%% long

% \bibliographystyle{/usr/share/texmf/tex/latex/texinput.RH6/kluwer/klunum}            %%% long

%\bibliographystyle{./PTCPmin}
 \bibliographystyle{PTCPmax}
 \bibliography{bib/JournAbbr,bib/Titov,bib/TitovLib,bib/Kaldor,bib/Isaev,bib/AbsConf/TitovAbs}

\newcommand{\noopsort}[1]{} \newcommand{\printfirst}[2]{#1}
  \newcommand{\singleletter}[1]{#1} \newcommand{\switchargs}[2]{#2#1}
\begin{thebibliography}{100}
\expandafter\ifx\csname url\endcsname\relax
  \def\url#1{{\tt #1}}\fi
\expandafter\ifx\csname urlprefix\endcsname\relax\def\urlprefix{URL }\fi
\providecommand{\eprint}[2][]{\url{#2}}

\bibitem{Erler:04}
Erler, J. \protect\BIBand{} {Ramsey-Musolf}, M.~J.
\newblock Low energy tests of the weak interaction (2004).
\newblock Eprint http:// arXiv.org/ hep-ph/0404291.

\bibitem{Glashow:61}
Glashow, S.~L.
\newblock Partial-symmetries of weak interactions.
\newblock {\em Nuclear Physics\/} {\bf 22}, 579--588 (1961).

\bibitem{Weinberg:67}
Weinberg, S.
\newblock A model of leptons.
\newblock {\em Phys.\ Rev.\ Lett.\/} {\bf 19}, 1264--1266 (1967).

\bibitem{Salam:68}
Salam, A.
\newblock Weak and electromagnetic interactions.
\newblock In: N.~Svartholm (ed.) {\em Elementary particle theory, relativistic
  groups, and analyticity\/}, pp. 367--377 (Almqvist and Wiksells, Stockholm,
  Sweden, 1968).

\bibitem{Weinberg:72}
Weinberg, S.
\newblock Effects of a neutral intermediate boson in semileptonic processes.
\newblock {\em Phys.\ Rev.\ D\/} {\bf 5}, 1412--1417 (1972).

\bibitem{Commins:99}
Commins, E.~D.
\newblock Electric dipole moments of leptons.
\newblock {\em Adv.\ At.\ Mol.\ Opt.\ Phys.\/} {\bf 40}, 1--55 (1999).

\bibitem{Christenson:64}
Christenson, J.~H., Cronin, J.~W., Fitch, V.~L., \protect\BIBand{} Turlay, R.
\newblock Evidence for the {$2\pi$} decay of the {$K_2^0$} meson.
\newblock {\em Phys.\ Rev.\ Lett.\/} {\bf 13}, 138--140 (1964).

\bibitem{Sapirstein:02aa}
Sapirstein, J.
\newblock Parity violation.
\newblock In: Schwerdtfeger  \cite{Schwerdtfeger:02bb}, pp. 468--522.

\bibitem{Berger:04}
Berger, R.
\newblock Parity-violation effects in molecules.
\newblock In: Schwerdtfeger  \cite{Schwerdtfeger:04aa}, pp. 188--288.

\bibitem{Ginges:04}
Ginges, J. S.~M. \protect\BIBand{} Flambaum, V.~V.
\newblock Violations of fundamental symmetries in atoms and tests of
  unification theories of elementary particles.
\newblock {\em Phys.\ Rep.\/} {\bf 397}, 63--154 (2004).

\bibitem{Landau:57}
Landau, L.~D.
\newblock Conservation laws in weak interactions.
\newblock {\em Sov.\ Phys.--JETP\/} {\bf 5}, 336--337 (1957).

\bibitem{Regan:02}
Regan, B.~C., Commins, E.~D., Schmidt, C.~J., \protect\BIBand{} DeMille, D.
\newblock New limit on the electron electric dipole moment.
\newblock {\em Phys.\ Rev.\ Lett.\/} {\bf 88}, 071805/1--4 (2002).

\bibitem{Romalis:01}
Romalis, M.~V., Griffith, W.~C., \protect\BIBand{} Fortson, E.~N.
\newblock New limit on the permanent electric dipole moment of {$^{199}$Hg}.
\newblock {\em Phys.\ Rev.\ Lett.\/} {\bf 86}, 2505 (2001).

\bibitem{Dmitriev:03}
Dmitriev, V.~F. \protect\BIBand{} {Sen'kov}, R.~A.
\newblock Schiff moment of the {M}ercury nucleus and the proton dipole moment.
\newblock {\em Phys.\ Rev.\ Lett.\/} {\bf 91}, 212303/1--4 (2003).

\bibitem{Sandars:67}
Sandars, P. G.~H.
\newblock Measurability of the proton electric dipole moment.
\newblock {\em Phys.\ Rev.\ Lett.\/} {\bf 19}, 1396--1398 (1967).

\bibitem{Sandars:65}
Sandars, P. G.~H.
\newblock The electric dipole moment of an atom.
\newblock {\em Phys.\ Lett.\/} {\bf 14}, 194--196 (1965).

\bibitem{Hinds:76}
Hinds, E.~A., Loving, C.~E., \protect\BIBand{} Sandars, P. G.~H.
\newblock Limits on {{\em P}} and {{\em T}} violating neutral current week
  interactions.
\newblock {\em Phys.\ Lett.\ B\/} {\bf 62}, 97--99 (1976).

\bibitem{Cho:91}
Cho, D., Sangster, K., \protect\BIBand{} Hinds, E.~A.
\newblock Search for time-reversal-symmetry violation in thallium fluoride
  using a jet source.
\newblock {\em Phys.\ Rev.\ A\/} {\bf 44}, 2783--2799 (1991).

\bibitem{Petrov:02}
Petrov, A.~N., Mosyagin, N.~S., Isaev, T.~A., Titov, A.~V., Ezhov, V.~F.,
  Eliav, E., \protect\BIBand{} Kaldor, U.
\newblock Calculation of the {P,T}-odd effects in {$^{205}$TlF} including
  electron correlation.
\newblock {\em Phys.\ Rev.\ Lett.\/} {\bf 88}, 073001/1--4 (2002).

\bibitem{Labzowsky:78}
Labzowsky, L.~N.
\newblock {$\Lambda$} doubling and parity nonconservation effects in the
  spectra of diatomic molecules.
\newblock {\em Sov.\ Phys.--JETP\/} {\bf 48}, 434--445 (1978).

\bibitem{Gorshkov:79}
Gorshkow, V.~G., Labzovsky, L.~N., \protect\BIBand{} Moskalyov, A.~N.
\newblock Space and time parity nonconservation effects in the spectra of
  diatomic molecules.
\newblock {\em Sov.\ Phys.--JETP\/} {\bf 49}, 209--216 (1979).

\bibitem{Sushkov:78}
Sushkov, O.~P. \protect\BIBand{} Flambaum, V.~V.
\newblock Parity violation effects in diatomic molecules.
\newblock {\em Sov.\ Phys.--JETP\/} {\bf 48}, 608--613 (1978).

\bibitem{Sushkov:84}
Sushkov, O.~P., Flambaum, V.~V., \protect\BIBand{} Khriplovich, I.~B.
\newblock To a possibility of investigation of {P}- and {T}-odd nuclear forces
  in atomic and molecular experiments.
\newblock {\em Sov.\ Phys.--JETP\/} {\bf 87}, 1521--1540 (1984).

\bibitem{Flambaum:85b}
Flambaum, V.~V. \protect\BIBand{} Khriplovich, I.~B.
\newblock On the enhancement of parity nonconserving effects in diatomic
  molecules.
\newblock {\em Phys.\ Lett.\ A\/} {\bf 110}, 121--125 (1985).

\bibitem{Kozlov:85}
Kozlov, M.~G.
\newblock Semiempirical calculation of {P}- and {P,T}-odd effects in diatomic
  molecules.
\newblock {\em Sov.\ Phys.--JETP\/} {\bf 62}, 1114 (1985).

\bibitem{Titov:85Dism}
Titov, A.~V.
\newblock {\it PhD Thesis}, {(St.-Petersburg (former Leningrad) State
  University, Russia, 1985)}.

\bibitem{Kozlov:87}
Kozlov, M.~G., Fomichev, V.~I., Dmitriev, Y.~Y., Labzovsky, L.~N.,
  \protect\BIBand{} Titov, A.~V.
\newblock Calculation of the {P}- and {T}-odd spin-ro\-ta\-ti\-onal
  {H}amil\-to\-ni\-an of the {PbF} molecule.
\newblock {\em J.\ Phys.\ B\/} {\bf 20}, 4939--4948 (1987).

\bibitem{Hudson:02}
Hudson, J.~J., Sauer, B.~E., Tarbutt, M.~R., \protect\BIBand{} Hinds, E.~A.
\newblock Measurement of the electron electic dipole moment using {YbF}
  molecules.
\newblock {\em Phys.\ Rev.\ Lett.\/} {\bf 89}, 023003/1--4 (2002).

\bibitem{DeMille:00}
DeMille, D., Bay, F., Bickman, S., Kawall, D., {Krause, Jr.}, D., Maxwell,
  S.~E., \protect\BIBand{} Hunter, L.~R.
\newblock Investigation of {PbO} as a system for measuring the electric dipole
  moment of the electron.
\newblock {\em Phys.\ Rev.\ A\/} {\bf 61}, 052507/1--8 (2000).

\bibitem{Okun:82}
{Okun'}, L.~B.
\newblock {\em Leptons and quarks\/} (North Holland, Amsterdam, 1982).

\bibitem{Kobayashi:73}
Kobayashi, M. \protect\BIBand{} Maskawa, T.
\newblock {CP}-violation in the renormalizable theory of weak interaction.
\newblock {\em Progr.\ Theor.\ Phys.\/} {\bf 49}, 652--657 (1973).

\bibitem{Kazakov:00}
Kazakov, D.~I.
\newblock Beyond the standard model ({I}n search of supersymmetry) (2000).
\newblock ArXiv: hep-ph/0012288.

\bibitem{Mohapatra:03}
Mohapatra, R.~N.
\newblock {\em Unification and Supersymmetry. {T}he Frontiers of Quark-Lepton
  Physics\/} (Springer, New-York, U.S.A., 2003).
\newblock 421\,pp.

\bibitem{Barr:92a}
Barr, S.~M.
\newblock Measurable {$T$-} and {$P$}-odd {$e{-}N$} interactions from {H}iggs
  boson exchange.
\newblock {\em Phys.\ Rev.\ Lett.\/} {\bf 68}, 1822--1825 (1992).

\bibitem{Barr:93b}
Barr, S.~M.
\newblock A review of {CP} violation in atoms.
\newblock {\em Int.\ J.\ Mod.\ Phys.\ A\/} {\bf 8}, 209--236 (1993).

\bibitem{Ginzburg:04}
Ginzburg, I.~F. \protect\BIBand{} Krawczyk, M.
\newblock Symmetries of two {H}iggs doublet model and {CP} violation (2004).
\newblock ArXiv: hep-ph/0408011.

\bibitem{Pati:74}
Pati, J.~C. \protect\BIBand{} Salam, A.
\newblock Lepton number as the fourth ``color''.
\newblock {\em Phys.\ Rev.\ D\/} {\bf 10}, 275--279 (1974).

\bibitem{Liu:94}
Liu, J.~T. \protect\BIBand{} Ng, D.
\newblock Lepton flavor changing processes and {CP} violation in the 331 model
  (1994).
\newblock ArXiv: hep-ph/9401228.

\bibitem{Masina:04}
Masina, I. \protect\BIBand{} Savoy, C.~A.
\newblock Changed lepton flavour and {CP} violation: {T}heoretical impact of
  present and future experiments (2004).
\newblock ArXiv: hep-ph/0410382.

\bibitem{Khriplovich:97}
Khriplovich, I.~B. \protect\BIBand{} Lamoraux, S.~K.
\newblock {\em {CP} Violation without Strangeness. The Electric Dipole Moments
  of Particles, Atoms, and Molecules\/} (Springer-Verlag, Berlin, 1997).

\bibitem{Schwerdtfeger:02bb}
Schwerdtfeger, P. (ed.).
\newblock {\em Relativistic Electronic Structure Theory. {Part~I}.
  {F}undamentals\/}, vol.~11 of {\em Theoretical and Computational Chemistry\/}
  (Elsevier, Amsterdam, 2002).
\newblock {xx\,+\,926\,pp}.

\bibitem{Schwerdtfeger:04aa}
Schwerdtfeger, P. (ed.).
\newblock {\em Relativistic Electronic Structure Theory. {P}art~2.
  {A}pplications\/}, vol.~14 of {\em Theoretical and Computational Chemistry\/}
  (Elsevier, Amsterdam, 2004).
\newblock {xv\,+\,787\,pp}.

\bibitem{Hirao:04}
Hirao, K. \protect\BIBand{} Ishikawa, Y. (eds.).
\newblock {\em Recent Advances in Relativistic Molecular Theory\/} (World
  Scientific, Singapore, 2004).
\newblock {328\,pp}.

\bibitem{Mosyagin:05a}
Mosyagin, N.~S., Petrov, A.~N., Titov, A.~V., \protect\BIBand{} Tupitsyn, I.~I.
\newblock {GRECPs} accounting for {B}reit effects in uranium, plutonium and
  superheavy elements 112, 113, 114.
\newblock {\em Progr.\ Theor.\ Chem.\ Phys.\/}  (2005).
\newblock In press; http://arXiv.org/physics/0505207.

\bibitem{Wood:78}
Wood, J.~H. \protect\BIBand{} Boring, A.~M.
\newblock Improved {P}auli {H}amiltonian for local-potential problems.
\newblock {\em Phys.\ Rev.\ B\/} {\bf 18}, 2701--2711 (1978).

\bibitem{Barthelat:80}
Barthelat, J.~C., Pelissier, M., \protect\BIBand{} Durand, P.
\newblock Analytical relativistic self-consistent-field calculations for atoms.
\newblock {\em Phys.\ Rev.\ A\/} {\bf 21}, 1773--1785 (1980).

\bibitem{Lenthe:93}
{van Lenthe}, E., Baerends, E.~J., \protect\BIBand{} Snijders, J.~G.
\newblock Relativistic regular two-component {H}amiltonians.
\newblock {\em J.\ Chem.\ Phys.\/} {\bf 99}, 4597--4610 (1993).

\bibitem{Wolf:02}
Wolf, A., Reiher, M., \protect\BIBand{} Hess, B.~A.
\newblock Two-component methods and the generalized {D}ouglas-{K}roll
  transformation.
\newblock In: Schwerdtfeger  \cite{Schwerdtfeger:02bb}, pp. 622--663.

\bibitem{Kutzelnigg:90}
Kutzelnigg, W.
\newblock Perturbation theory of relativistic corrections. 2. {A}nalysis and
  classification of known and other possible methods.
\newblock {\em Z.\ Phys.\ D\/} {\bf 15}, 27--50 (1990).

\bibitem{Dyall:02a}
Dyall, K.~G.
\newblock A systematic sequence of relativistic approximations.
\newblock {\em J.\ Comput.\ Chem.\/} {\bf 23}, 786--793 (2002).

\bibitem{Visscher:02aa}
Visscher, L.
\newblock The {D}irac equation in quantum chemistry: {S}trategies to overcome
  the current computational problems.
\newblock {\em J.\ Comput.\ Chem.\/} {\bf 23}, 759--766 (2002).

\bibitem{Grant:04A}
Grant, I.~P. \protect\BIBand{} Quiney, H.
\newblock {BERTHA} -- relativistic atomic/molecular structure (Les Houches,
  France, 2004).
\newblock Conference ``Quantum Systems in Chemistry and Physics: {QSCP-IX}'',
  oral and poster reports.

\bibitem{DIRAC}
Saue, T., Jensen, H. J.~A., Visscher, L., {\em et~al\/}.
\newblock {``{\sc dirac04}''} (2004).
\newblock A relativistic {\em ab~initio} electronic structure program.

\bibitem{Quiney:99}
Quiney, H.~M., Skaane, H., \protect\BIBand{} Grant, I.~P.
\newblock Ab initio relativistic quantum chemistry: four-components good,
  two-components bad!
\newblock {\em Adv.\ Quantum Chem.\/} {\bf 32}, 1--49 (1999).

\bibitem{BERTHA}
Grant, I.~P., Quiney, H.~M., \protect\BIBand{} Skaane, H.
\newblock {``{\sc bertha}''} (1998).
\newblock An ab~initio relativistic molecular electronic structure program
  \cite{Quiney:99}.

\bibitem{Lee:77}
Lee, Y.~S., Ermler, W.~C., \protect\BIBand{} Pitzer, K.~S.
\newblock Ab initio effective core potentials including relativistic effects.
  {I}. {F}ormalism and applications to the {Xe} and {Au} atoms.
\newblock {\em J.\ Chem.\ Phys.\/} {\bf 67}, 5861--5876 (1977).

\bibitem{Ermler:88}
Ermler, W.~C., Ross, R.~B., \protect\BIBand{} Christiansen, P.~A.
\newblock Spin-orbit coupling and other relativistic effects in atoms and
  molecules.
\newblock {\em Adv.\ Quantum Chem.\/} {\bf 19}, 139--182 (1988).

\bibitem{Schwerdtfeger:03}
Schwerdtfeger, P.
\newblock Relativistic pseudopotentials.
\newblock In: U.~Kaldor \protect\BIBand{} S.~Wilson (eds.) {\em Theoretical
  chemistry and physics of heavy and superheavy elements\/}, pp. 399--438
  (Kluwer academic publishers, Dordrecht, The Netherlands, 2003).

\bibitem{YSLee:04}
Lee, Y.~S.
\newblock Two-component relativistic effective core potential calculations for
  molecules.
\newblock In: Schwerdtfeger  \cite{Schwerdtfeger:04aa}, pp. 352--416.

\bibitem{Teichteil:04}
Teichteil, C., Maron, L., \protect\BIBand{} Vallet, V.
\newblock Relativistic pseudopotential calculations for electronic excited
  states.
\newblock In: Schwerdtfeger  \cite{Schwerdtfeger:04aa}, pp. 476--551.

\bibitem{Blochl:90}
{Bl\"ochl}, P.~E.
\newblock Generalized separable potentials for electronic-structure
  calculations.
\newblock {\em Phys.\ Rev.\ B\/} {\bf 41}, 5414--5416 (1990).

\bibitem{Vanderbilt:90}
Vanderbilt, D.
\newblock Soft self-consistent pseudopotentials in a generalized eigenvalue
  formalism.
\newblock {\em Phys.\ Rev.\ B\/} {\bf 41}, 7892--7895 (1990).

\bibitem{Theurich:01}
Theurich, G. \protect\BIBand{} Hill, N.~A.
\newblock Self-consistent treatment of spin-orbit coupling in solids using
  relativistic fully separable {\em ab initio} pseudopotentials.
\newblock {\em Phys.\ Rev.\ B\/} {\bf 64}, 073106, 1--4 (2001).

\bibitem{Bonifacic:74}
Bonifacic, V. \protect\BIBand{} Huzinaga, S.
\newblock Atomic and molecular calculations with the model potential method.
  {I}.
\newblock {\em J.\ Chem.\ Phys.\/} {\bf 60}, 2779--2786 (1974).

\bibitem{Katsuki:88b}
Katsuki, S. \protect\BIBand{} Huzinaga, S.
\newblock An effective {H}amiltonian method for valence-electron molecular
  calculations.
\newblock {\em Chem.\ Phys.\ Lett.\/} {\bf 152}, 203--206 (1988).

\bibitem{Seijo:04}
Seijo, L. \protect\BIBand{} Barandiar{\'a}n, Z.
\newblock Relativistic ab-initio model potential calculations for molecules and
  embedded clusters.
\newblock In: Schwerdtfeger  \cite{Schwerdtfeger:04aa}, pp. 417--475.

\bibitem{Petrov:04b}
Petrov, A.~N., Mosyagin, N.~S., Titov, A.~V., \protect\BIBand{} Tupitsyn, I.~I.
\newblock Accounting for the {B}reit interaction in relativistic effective core
  potential calculations of actinides.
\newblock {\em J.\ Phys.\ B\/} {\bf 37}, 4621--4637 (2004).

\bibitem{Dyall:94}
Dyall, K.~G.
\newblock An exact separation of the spin-free and spin-dependent terms of the
  {D}irac-{C}oulomb-{B}reit {H}amiltonian.
\newblock {\em J.\ Chem.\ Phys.\/} {\bf 100}, 2118--2127 (1994).

\bibitem{Titov:99}
Titov, A.~V. \protect\BIBand{} Mosyagin, N.~S.
\newblock Generalized relativistic effective core potential: {T}heoretical
  grounds.
\newblock {\em Int.\ J.\ Quantum Chem.\/} {\bf 71}, 359--401 (1999).

\bibitem{Mosyagin:04a}
Mosyagin, N.~S. \protect\BIBand{} Titov, A.~V.
\newblock Accounting for correlations with core electrons by means of the
  generalized {RECP}: Atoms {Hg} and {Pb} and their compounds.
\newblock ArXiv.org/ physics/0406143; J.\ Chem.\ Phys., v.122 (2005).

\bibitem{Durand:75}
Durand, P. \protect\BIBand{} Barthelat, J.-C.
\newblock A theoretical method to determine atomic pseudopotentials for
  electronic structure calculations of molecules and solids.
\newblock {\em Theor.\ Chim.\ Acta\/} {\bf 38}, 283--302 (1975).

\bibitem{Christiansen:79}
Christiansen, P.~A., Lee, Y.~S., \protect\BIBand{} Pitzer, K.~S.
\newblock Improved {\em ab~initio} effective core potentials for molecular
  calculations.
\newblock {\em J.\ Chem.\ Phys.\/} {\bf 71}, 4445--4450 (1979).

\bibitem{Hamann:79}
Hamann, D.~R., {Schl\"uter}, M., \protect\BIBand{} Chiang, C.
\newblock Norm-conserving pseudopotentials.
\newblock {\em Phys.\ Rev.\ Lett.\/} {\bf 43}, 1494--1497 (1979).

\bibitem{Titov:00a}
Titov, A.~V. \protect\BIBand{} Mosyagin, N.~S.
\newblock The generalized relativistic effective core potential method:
  {T}heory and calculations.
\newblock {\em Russ.\ J.\ Phys.\ Chem.\/} {\bf 74, {\rm Suppl.\,2}}, S376--387
  (2000).
\newblock [arXiv: physics/0008160].

\bibitem{Titov:02Dism}
Titov, A.~V.
\newblock {\it Doctorate Thesis}, {(Petersburg Nuclear Physics Institute, RAS,
  Russia, 2002)}.

\bibitem{Phillips:59}
Phillips, J.~C. \protect\BIBand{} Kleinman, L.
\newblock New method for calculating wave functions in crystals and molecules.
\newblock {\em Phys.\ Rev.\/} {\bf 116}, 287--294 (1959).

\bibitem{Pacios:85}
Pacios, L.~F. \protect\BIBand{} Christiansen, P.~A.
\newblock Ab initio relativistic effective potentials with spin-orbit
  operators. {I}. {Li} through {Ag}.
\newblock {\em J.\ Chem.\ Phys.\/} {\bf 82}, 2664--2671 (1985).

\bibitem{Blochl:94}
{Bl\"ochl}, P.~E.
\newblock Projector augmented-wave method.
\newblock {\em Phys.\ Rev.\ B\/} {\bf 50}, 17953--17979 (1994).

\bibitem{Titov:92A}
Titov, A.~V.
\newblock A restoration of wave function in core region of a molecule.
\newblock In: {\em Theses of reports of the 4th European Conf.\ on Atomic and
  Mol.\ Physics\/}, p. 299 (Riga, Latvia, 1992).

\bibitem{Titov:96}
Titov, A.~V.
\newblock A two-step method of calculation of the electronic structure of
  molecules with heavy atoms: {T}heoretical aspect.
\newblock {\em Int.\ J.\ Quantum Chem.\/} {\bf 57}, 453--463 (1996).

\bibitem{Desclaux:74b}
Desclaux, J.~P. \protect\BIBand{} {Pyykk\"o}, P.
\newblock Relativistic and non-relativistic {H}artree-{F}ock one-centre
  expansion calculations for the series {CH$_{4}$ to PbH$_{4}$} within the
  spherical approximation.
\newblock {\em Chem.\ Phys.\ Lett.\/} {\bf 29}, 534 (1974).

\bibitem{Desclaux:76c}
Desclaux, J.~P. \protect\BIBand{} {Pyykk\"o}, P.
\newblock {D}irac-{F}ock one-centre calculations. {T}he molecules {CuH, AgH}
  and {AuH} including $p$-type symmetry functions.
\newblock {\em Chem.\ Phys.\ Lett.\/} {\bf 39}, 300--303 (1976).

\bibitem{Desclaux:02}
Desclaux, J.-P.
\newblock Tour historique.
\newblock In: Schwerdtfeger  \cite{Schwerdtfeger:02bb}, pp. 1--22.

\bibitem{Pitzer:79}
Pitzer, K.~S.
\newblock Relativistic effects on chemical properties.
\newblock {\em Acc.\ Chem.\ Res.\/} {\bf 12}, 271--276 (1979).

\bibitem{Pyykko:79}
{Pyykk\"o}, P. \protect\BIBand{} Desclaux, J.-P.
\newblock Relativity and the periodic system of elements.
\newblock {\em Acc.\ Chem.\ Res.\/} {\bf 12}, 276--281 (1979).

\bibitem{Hinds:80a}
Hinds, E.~A. \protect\BIBand{} Sandars, P. G.~H.
\newblock Electric dipole hyperfine structure of {TlF}.
\newblock {\em Phys.\ Rev.\ A\/} {\bf 21}, 471--479 (1980).

\bibitem{Coveney:83}
Coveney, P.~V. \protect\BIBand{} Sandars, P. G.~H.
\newblock Parity- and time-violating interactions in thallium fluoride.
\newblock {\em J.\ Phys.\ B\/} {\bf 16}, 3727--3740 (1983).

\bibitem{Laerdahl:97}
Laerdahl, J.~K., Saue, T., {Faegri, Jr}, K., \protect\BIBand{} Quiney, H.~M.
\newblock Ab~initio study of {PT}-odd interactions in {T}hallium {F}luoride.
\newblock {\em Phys.\ Rev.\ Lett.\/} {\bf 79}, 1642--1645 (1997).

\bibitem{Parpia:97}
Parpia, F.~A.
\newblock Electric-dipole hyperfine matrix elements of the ground state of the
  {TlF} molecule in the {D}irac-{F}ock approximation.
\newblock {\em J.\ Phys.\ B\/} {\bf 30}, 3983--4001 (1997).

\bibitem{Dmitriev:92}
Dmitriev, Y.~Y., Khait, Y.~G., Kozlov, M.~G., Labzovsky, L.~N., Mitrushenkov,
  A.~O., Shtoff, A.~V., \protect\BIBand{} Titov, A.~V.
\newblock Calculation of the spin-ro\-ta\-ti\-onal hamiltonian including {P}-
  and {P,T}-odd weak interaction terms for the {HgF} and {PbF} molecules.
\newblock {\em Phys.\ Lett.\ A\/} {\bf 167}, 280--286 (1992).

\bibitem{Titov:92}
Titov, A.~V.
\newblock Matrix elements of the {$U(2n)$} generators in the spin-or\-bit
  basis.
\newblock {\em Int.\ J.\ Quantum Chem.\/} {\bf 42}, 1711--1716 (1992).

\bibitem{Titov:96b}
Titov, A.~V., Mosyagin, N.~S., \protect\BIBand{} Ezhov, V.~F.
\newblock {$P,T$}-odd spin-rotational {H}amiltonian for the {YbF} molecule.
\newblock {\em Phys.\ Rev.\ Lett.\/} {\bf 77}, 5346--5349 (1996).

\bibitem{Kozlov:97}
Kozlov, M.~G., Titov, A.~V., Mosyagin, N.~S., \protect\BIBand{} Souchko, P.~V.
\newblock Enhancement of the electric dipole moment of the electron in the
  {BaF} molecule.
\newblock {\em Phys.\ Rev.\ A\/} {\bf 56}, R3326--3329 (1997).

\bibitem{Mosyagin:98}
Mosyagin, N.~S., Kozlov, M.~G., \protect\BIBand{} Titov, A.~V.
\newblock Electric dipole moment of the electron in the {YbF} molecule.
\newblock {\em J.\ Phys.\ B\/} {\bf 31}, L763--767 (1998).

\bibitem{Quiney:98}
Quiney, H.~M., Skaane, H., \protect\BIBand{} Grant, I.~P.
\newblock Hyperfine and {P,T}-odd effects in {YbF} {$^2{\Sigma}$}.
\newblock {\em J.\ Phys.\ B\/} {\bf 31}, L85--95 (1998).

\bibitem{Parpia:98}
Parpia, F.
\newblock {\em Ab initio} calculation of the enhancement of the {EDM} of an
  electron in the {YbF}.
\newblock {\em J.\ Phys.\ B\/} {\bf 31}, 1409--1430 (1998).

\bibitem{Isaev:04}
Isaev, T.~A., Petrov, A.~N., Mosyagin, N.~S., Titov, A.~V., Eliav, E.,
  \protect\BIBand{} Kaldor, U.
\newblock In search of the electron dipole moment: {\it Ab initio} calculations
  on {$^{207}$PbO} excited states.
\newblock {\em Phys.\ Rev.\ A\/} {\bf 69}, 030501(R)/1--4 (2004).

\bibitem{Petrov:05a}
Petrov, A.~N., Titov, A.~V., Isaev, T.~A., Mosyagin, N.~S., \protect\BIBand{}
  DeMille, D.~P.
\newblock Configuration interaction calculation of hyperfine and {P,T}-odd
  constants on {$^{207}$PbO} excited states for the electron {EDM}.
\newblock {\em Phys.\ Rev.\ A\/}  (2005).
\newblock In press; http://arXiv.org/physics/0409045.

\bibitem{Isaev:04b}
Isaev, T.~A., Mosyagin, N.~S., Petrov, A.~N., \protect\BIBand{} Titov, A.~V.
\newblock On the electron {EDM} enhancement in {HI$^+$} (2004).
\newblock ArXiv: physics/0412177, 3\,p.

\bibitem{Dzuba:96}
Dzuba, V.~A., Flambaum, V.~V., \protect\BIBand{} Kozlov, M.~G.
\newblock Calculation of energy levels for atoms with several valence
  electrons.
\newblock {\em JETP Lett.\/} {\bf 63}, 882--887 (1996).

\bibitem{Kaldor:97}
Kaldor, U.
\newblock Relativistic coupled cluster: {M}ethod and applications.
\newblock In: R.~J. Bartlett (ed.) {\em Recent Advances in Coupled-Cluster
  Methods\/}, pp. 125--153 (World Scientific, Singapore, 1997).

\bibitem{Landau:01c}
Landau, A., Eliav, E., \protect\BIBand{} Kaldor, U.
\newblock Intermediate {H}amiltonian {F}ock-space coupled-cluster method.
\newblock {\em Adv.\ Quantum Chem.\/} {\bf 39}, 171--188 (2001).

\bibitem{Buenker:99}
Buenker, R.~J. \protect\BIBand{} Krebs, S.
\newblock The configuration-driven approach for multireference configuration
  interaction calculations.
\newblock In: K.~Hirao (ed.) {\em Recent Advances in Multireference Methods\/},
  pp. 1--29 (World Scientific, Singapore, 1999).

\bibitem{Alekseyev:04}
Alekseyev, A.~B., Liebermann, H.-P., \protect\BIBand{} Buenker, R.~J.
\newblock Spin-orbit multireference configuration interaction method and
  applications to systems containing heavy atoms.
\newblock In: Hirao \protect\BIBand{} Ishikawa  \cite{Hirao:04}, pp. 65--105.

\bibitem{Titov:01}
Titov, A.~V., Mosyagin, N.~S., Alekseyev, A.~B., \protect\BIBand{} Buenker,
  R.~J.
\newblock {GRECP/MRD-CI} calculations of spin-orbit splitting in ground state
  of {Tl} and of spectroscopic properties of {TlH}.
\newblock {\em Int.\ J.\ Quantum Chem.\/} {\bf 81}, 409--421 (2001).

\bibitem{Dolg:00aa}
Dolg, M.
\newblock Effective core potentials.
\newblock In: J.~Grotendorst (ed.) {\em Modern Methods and Algorithms of
  Quantum Chemistry\/}, vol.~1 of {\em NIC Series\/}, pp. 479--508 (J{\"u}lich,
  2000).
\newblock {[http://www.fz-juelich.de]}.

\bibitem{Titov:00bmin}
Titov, A.~V. \protect\BIBand{} Mosyagin, N.~S.
\newblock Comments on {``Effective Core Potentials'' by M.Dolg}
  \cite{Dolg:00aa}.
\newblock ArXiv.org/ physics/0008239 (2000).

\bibitem{Titov:95}
Titov, A.~V. \protect\BIBand{} Mosyagin, N.~S.
\newblock Self-con\-sis\-tent relativistic effective core potentials for
  transition metal atoms: {Cu, Ag}, and {Au}.
\newblock {\em Structural Chem.\/} {\bf 6}, 317--321 (1995).

\bibitem{HFDB}
Tupitsyn, I.~I.
\newblock {``{\sc hfdb}''} (2003).
\newblock Program for atomic finite-difference four-component
  {D}irac-{H}artree-{F}ock-{B}reit calculations written on the base of the {\sc
  hfd} code~\cite{Bratzev:77}.

\bibitem{Bratzev:77}
Bratzev, V.~F., Deyneka, G.~B., \protect\BIBand{} Tupitsyn, I.~I.
\newblock Application of the {H}artree-{F}ock method to calculation of
  relativistic atomic wave functions.
\newblock {\em Bull.\ Acad.\ Sci.\ USSR, Phys.\ Ser.\/} {\bf 41}, 173--182
  (1977).

\bibitem{Tupitsyn:02A}
Tupitsyn, I.~I. \protect\BIBand{} Petrov, A.~N.
\newblock Self--consistent field calculations of heavy atoms including the
  breit interaction.
\newblock In: {\em 5--th Session of the V.A. Fock School on Quantum and
  Computational Chemistry\/}, p.~62 (Novgorod the Great, 2002).

\bibitem{HFJ}
Tupitsyn, I.~I. \protect\BIBand{} Mosyagin, N.~S.
\newblock {``{\sc grecp/hfj}''} (1995).
\newblock Program for atomic finite-difference two-component {H}artree-{F}ock
  calculations with the {g}eneralized {RECP} in the {$jj$}-coupling scheme.

\bibitem{Tupitsyn:95}
Tupitsyn, I.~I., Mosyagin, N.~S., \protect\BIBand{} Titov, A.~V.
\newblock Generalized relativistic effective core potential. {I.
  N}u\-me\-ri\-cal calculations for atoms {Hg} through {Bi}.
\newblock {\em J.\ Chem.\ Phys.\/} {\bf 103}, 6548--6555 (1995).

\bibitem{Lindgren:84}
Lindgren, I.
\newblock Effective operators in the atomic hyperfine interaction.
\newblock {\em Rep.\ Prog.\ Phys.\/} {\bf 47}, 345--398 (1984).

\bibitem{Kunik:71}
Kunik, D. \protect\BIBand{} Kaldor, U.
\newblock Hyperfine pressure shift of hydrogen in helium.
\newblock {\em J.\ Chem.\ Phys.\/} {\bf 55}, 4127--4131 (1971).

\bibitem{Monkhorst:77}
Monkhorst, H.~J.
\newblock Calculation of properties with the coupled-cluster method.
\newblock {\em Int.\ J.\ Quantum Chem.: Quantum Chem.\ Symp.\/} {\bf 11},
  421--432 (1977).

\bibitem{Kozlov:95}
Kozlov, M. \protect\BIBand{} Labzowsky, L.
\newblock Parity violation effects in diatomics.
\newblock {\em J.\ Phys.\ B\/} {\bf 28}, 1931--1961 (1995).

\bibitem{Knight:81}
{Knight, Jr.}, L.~B., Fisher, T.~A., \protect\BIBand{} Wise, M.~B.
\newblock Photolytic codeposition generation of the {HgF} radical in an argon
  matrix at 12 {K: An ESR} investigation.
\newblock {\em J.\ Chem.\ Phys.\/} {\bf 74}, 6009--6013 (1981).

\bibitem{Huber:79}
Huber, K.~P. \protect\BIBand{} Herzberg, G.
\newblock {\em Constants of Diatomic Molecules\/} (Van Nostrand-Reinhold, New
  York, 1979).

\bibitem{Mosyagin:XXa}
Mosyagin, N.~S. {\em et~al\/}.
\newblock {GRECP/NOCR} calculations of hyperfine structure and parity violation
  effects in {YbF}, {HgF} and {HgH} (2002).
\newblock Unpublished.

\bibitem{Kozlov:94}
Kozlov, M.~G. \protect\BIBand{} Ezhov, V.~F.
\newblock Enhancement of the electric dipole moment of the electron in the
  {YbF} molecule.
\newblock {\em Phys.\ Rev.\ A\/} {\bf 49}, 4502--4506 (1994).

\bibitem{Knight:70}
{Knight, Jr.}, L.~B. \protect\BIBand{} {Weltner, Jr.}, W.
\newblock On the spin-doubling constant, {$\gamma$}, and {$g$} tensor in
  {$^2\Sigma$} molecules.
\newblock {\em J.\ Chem.\ Phys.\/} {\bf 53}, L4111--4112 (1970).

\bibitem{Kozlov:97c}
Kozlov, M.~G.
\newblock Enhancement of the electric dipole moment of the electron in the
  {YbF} molecule.
\newblock {\em J.\ Phys.\ B\/} {\bf 30}, L607--612 (1997).

\bibitem{Knight:71}
Knight, L.~B., Easley, W.~C., \protect\BIBand{} Weltner, W.
\newblock Hyperfine interaction and chemical bonding in {MgF, CaF}, {SrF}, and
  {BaF} molecules.
\newblock {\em J.\ Chem.\ Phys.\/} {\bf 54}, 322--329 (1971).

\bibitem{Ryzlewicz:82}
Ryzlewicz, C., {Sch\"utze-Pahlmann}, H.~U., Hoeft, J., \protect\BIBand{}
  {T\"orring}, T.
\newblock Rotational spectrum and hyperfine structure of the {$^2\Sigma$}
  radicals {BaF} and {BaCl}.
\newblock {\em Chem.\ Phys.\/} {\bf 71}, 389--399 (1982).

\bibitem{Schiff:63}
Schiff, L.~I.
\newblock Measurability of nuclear electric dipole moments.
\newblock {\em Phys.\ Rev.\/} {\bf 132}, 2194--2200 (1963).

\bibitem{Quiney:98b}
Quiney, H.~M., Laerdahl, J.~K., {Faegri, Jr}, K., \protect\BIBand{} Saue, T.
\newblock Ab initio {D}irac-{H}artree-{F}ock calculations of chemical
  properties and {PT}-odd effects in thallium fluoride.
\newblock {\em Phys.\ Rev.\ A\/} {\bf 57}, 920--944 (1998).

\bibitem{Wilson:94}
Wilson, S., Moncrieff, D., \protect\BIBand{} Kobus, J.
\newblock {TlF $(^1\Sigma^+)$}: Some preliminary electronic structure
  calculations (1994).
\newblock RAL-94-082 Report.

\bibitem{Dzuba:02}
Dzuba, V.~A., Flambaum, V.~V., Ginges, J. S.~M., \protect\BIBand{} Kozlov,
  M.~G.
\newblock Electric dipole moments of {Hg, Xe, Rn, Ra, Pu}, and {TlF} induced by
  the nuclear schiff moment and limits on time-reversal violating interactions.
\newblock {\em Phys.\ Rev.\ A\/} {\bf 66}, 012111 (2002).

\bibitem{Kaldor:99}
Kaldor, U. \protect\BIBand{} Eliav, E.
\newblock High-accuracy calculations for heavy and super-heavy elements.
\newblock {\em Adv.\ Quantum Chem.\/} {\bf 31}, 313--336 (1999).

\bibitem{Kaldor:04b}
Kaldor, U., Eliav, E., \protect\BIBand{} Landau, A.
\newblock Four-component relativistic coupled cluster -- method and
  applications.
\newblock In: Hirao \protect\BIBand{} Ishikawa  \cite{Hirao:04}, pp. 283--327.

\bibitem{Paldus:99}
Paldus, J. \protect\BIBand{} Li, X.
\newblock A critical assessment of coupled cluster method in quantum chemistry.
\newblock {\em Adv.\ Chem.\ Phys.\/} {\bf 110}, 1--175 (1999).

\bibitem{Paldus:03}
Paldus, J.
\newblock Coupled cluster methods.
\newblock In: S.~Wilson (ed.) {\em Handbook of Molecular Physics and Quantum
  Chemistry\/}, vol.~2, pp. 272--313 (John Wiley \& Sons, Ltd, Chichester,
  2003).

\bibitem{Egorov:01}
Egorov, D., Weinstein, J.~D., Patterson, D., Friedrich, B., \protect\BIBand{}
  Doyle, J.~M.
\newblock Spectroscopy of laser-ablated buffer-gas-cooled {PbO} at 4\,{K} and
  the prospects for measuring the electric dipole moment of the electron.
\newblock {\em Phys.\ Rev.\ A\/} {\bf 63}, 030501(R)/1--4 (2001).

\bibitem{Buenker:74}
Buenker, R.~J. \protect\BIBand{} Peyerimhoff, S.~D.
\newblock Individualized configuration selection in {CI} calculations with
  subsequent energy extrapolation.
\newblock {\em Theor.\ Chim.\ Acta\/} {\bf 35}, 33--58 (1974).

\bibitem{Isaev:00}
Isaev, T.~A., Mosyagin, N.~S., Kozlov, M.~G., Titov, A.~V., Eliav, E.,
  \protect\BIBand{} Kaldor, U.
\newblock Accuracy of {RCC-SD} and {PT2/CI} methods in all-electron and {RECP}
  calculations on {Pb} and {Pb$^{2+}$}.
\newblock {\em J.\ Phys.\ B\/} {\bf 33}, 5139--5149 (2000).

\bibitem{Martin:88}
Martin, F., Bacis, R., Verges, J., Bachar, J., \protect\BIBand{} Rosenwaks, S.
\newblock High resolution fourier transform spectroscopy of {PbO} molecule from
  investigation of the {O$_2(^1Dg)-$Pb} reaction.
\newblock {\em Spectrochim.\ Acta\/} {\bf 44A}, 889 (1988).

\bibitem{Hunter:02}
Hunter, L.~R., Maxwell, S.~E., Ulmer, K.~A., Charney, N.~D., Peck, S.~K.,
  Krause, D., {Ter-Avetisyan}, S., \protect\BIBand{} DeMille, D.
\newblock Precise spectroscopy of the {$a(1)~[^3\Sigma^+]$} state of {PbO}.
\newblock {\em Phys.\ Rev.\ A\/} {\bf 65}, 030501(R) (2002).

\bibitem{Kawall:04a}
Kawall, D., Gurevich, Y., \protect\BIBand{} DeMille, D.
\newblock To be published.

\end{thebibliography}

%============================================================================
% Tables:

%============================================================================
%\include{apdx}       %%%     List of abbreviations

%\printindex

\end{document}